\documentclass[aps,prb,superscriptaddress,amsfonts,amsmath,amssymb,twocolumn,showpacs,floatfix,footinbib]{revtex4-1}

\usepackage{amsmath}
\usepackage{amssymb}
\usepackage{graphicx}

\newcommand{\beq}{\begin{equation}}
\newcommand{\eeq}{\end{equation}}
\newcommand{\bma}{\begin{math}}
\newcommand{\ema}{\end{math}}
\newcommand{\beqa}{\begin{eqnarray}}
\newcommand{\eeqa}{\end{eqnarray}}

\newcommand{\id}{{\mathbf 1}}

\allowdisplaybreaks[4]

\newcommand{\be}{\begin{equation}}
\newcommand{\ee}{\end{equation}}
\newcommand{\bq}{\begin{eqnarray}}
\newcommand{\eq}{\end{eqnarray}}

\newcommand{\rf}[1]{(\ref{#1})}

\newcommand{\ch}{{\rm ch}}

\begin{document}

\title{A hierarchy of exactly solvable spin-1/2 chains with $so(N)_1$ critical points}

\author{Ville Lahtinen}
\affiliation{%
Institute for Theoretical Physics, University of Amsterdam,
Science Park 904,
1090 GL Amsterdam,
The Netherlands
}
\affiliation{
Institute-Lorentz for Theoretical Physics, Leiden University,
PO Box 9506, NL-2300 RA Leiden, The Netherlands}

\author{Teresia M\r{a}nsson}
\affiliation{%
Department of Theoretical Physics, School of Engineering Sciences,
Royal Institute of Technology (KTH),
Roslagstullsbacken 21, SE-106 91 Stockholm, Sweden
}

\author{Eddy Ardonne}
\affiliation{%
Department of Physics, Stockholm University,
AlbaNova University Center, SE-106 91 Stockholm, Sweden
}

\date{\today}

\begin{abstract}
We construct a hierarchy of exactly solvable spin-1/2 chains with $so(N)_1$ critical points. Our construction is based on the framework of condensate-induced transitions between 
topological phases. We employ this framework to construct a Hamiltonian term that couples $N$ transverse field Ising chains such that the resulting theory is critical and described 
by the $so(N)_1$ conformal field theory. By employing spin duality transformations, we then cast these spin chains for arbitrary $N$ into translationally invariant forms that all allow exact solution by the means of a Jordan-Wigner transformation. For odd $N$ our models generalize the phase diagram of the transverse field Ising chain, the simplest model in our hierarchy. For even $N$ the models can be viewed as longer ranger generalizations of the XY chain, the next model in the hierarchy. We also demonstrate that our method of constructing spin chains with given critical points goes beyond exactly solvable models. Applying the same strategy to the Blume-Capel model, a spin-1 generalization of the Ising chain in a generic magnetic field, we construct another critical spin-1 chain with the predicted CFT describing the criticality.
\end{abstract}


\maketitle

\section{Introduction}

Spin chains have played a crucial role in the study of magnetism
ever since Bethe's solution of the Heisenberg spin-chain \cite{bethe31}.
As relatively simple models that exhibit distinct phases of matter, they have also tremendously contributed to our understanding of quantum phase transitions\cite{book:sachdev}.
 In particular, the Ising spin chain with a transverse magnetic field has been
the prototype model to learn about quantum criticality. The exact solution\cite{epw70,p70}
by means of a Jordan-Wigner transformation\cite{lsm61} has enabled one to
study its properties in great detail, especially at its critical point.
Critical points, be them thermal or quantum, are particularly interesting, because they display universal
behavior. Quantum critical points in one-dimensional systems are often described by a conformal field theory (CFT)\cite{bpz84,byb}, which encodes the spectral structure and the characteristic algebraic decay of correlation functions. Remarkably, the description by a CFT is not limited strictly to the critical point, but it also allows one to calculate most of the physical properties of the model in its vicinity.

For the understanding of quantum matter, it is therefore desirable to have (preferably exactly solvable, if possible) models that exhibit interesting critical points described by different CFTs. 
Some examples are already known. For instance, the (quantum) phase transition of the transverse field
Ising chain is described by the Ising CFT and is said to belong to the Ising universality class, as the same CFT describes the critical point of the
classical two-dimensional Ising model. Other classic examples of integrability are the spin-1/2 XY chain, whose criticality is described by the so called $u(1)_4$ CFT, and the spin-1 Heisenberg model, whose solution by means of a (nested) Bethe Ansatz\cite{zamolodchkov80,takhtajan82,babujian83} has enabled to confirm the existence of an $su(2)_2$ critical point. The desirable exact solvability is a scarce property though and in general restricted to special spin chains with nearest-neighbor interactions. A rare example of an exactly solvable model with long range interactions is the celebrated spin-1/2 Haldane-Shastry model \cite{haldane88,shastry88}. Remarkably, this model can be generalized to a series of critical spin-$S$ chains with an $su(2)_{k=2S}$ CFT 
description\cite{nielsen11,book:greiter}. Recent studies\cite{michaud13} suggest that this same series of critical points can also appear with local interactions in the spin-$S$ generalizations of the Majundar-Gosh spin chain\cite{mg69a,mg69b}. Another recent approach for exact solvability in higher spin systems has been to start with manifestly $so(N)$ symmetric spin chains.\cite{tu08} Several subsequent case studies strongly support the conjecture that such models exhibit critical points described by the $so(N)_1$ CFTs.\cite{alet11,tu11,orus11,capponi13}

While examples exist, it would be desirable to have a systematic framework for exactly solvable models with interesting critical points. However, this is a non-trivial task for a simple reason.
While it is easy to verify that a critical 
point is described by a given CFT, there is in general no simple prescription to write down spin chains with critical points described by chosen CFTs. Recently, few possible routes around this
 have been pointed out by borrowing ideas from two dimensional topologically ordered systems. In Ref.~\onlinecite{Tu13} a series of long-range parent Hamiltonians were derived from trial 
wave functions, that enabled to numerically verify the existence of  critical points described by the $so(N)_1$ CFTs. Another route was pointed out by us in Ref.~\onlinecite{Mansson13}, where we
argued that a class of phase transitions between topologically ordered phases -- the so called condensate-induced transitions\cite{Bais09} -- could be used to derive spin chains with given 
critical points. In a nutshell, this argument is based on the close connection between two-dimensional gapped topological phases and one-dimensional gapless systems\cite{Witten89}. For a 
large class of topological phases the topological quantum field theory describing the (anyonic) quasiparticle excitations of the gapped bulk is in one-to-one correspondence with the CFT 
describing its gapless edge. We argued that if two topologically phases are related by a condensation transition, then also two critical spin chains described by the respective CFTs 
should be related.

In this paper, we fully exploit this insight and construct a general hierarchy of exactly solvable spin-1/2 chains with $so(N)_1$ critical points. While for generic $N$ these chains contain $N$-spin 
interaction terms, their form is such that every model can be solved
straightforwardly by means of a Jordan-Wigner transformation.\cite{suzuki71b} The two simplest models in our hierarchy, the cases $N=1$ and $N=2$,
are the well known transverse field Ising model (TFI) and the XY model, respectively. Of the other models, the physically most interesting model and one of our main results is the case of $N=3$. It exhibits $so(3)_1 \simeq su(2)_2$ criticality that has not been previously discovered in local and exactly solvable spin-1/2 models. Explicitly, at criticality this spin chain takes the form
\begin{align}
\label{H3_first}
&H_{su(2)_2} = 
\sum_j
\bigl(  \tau^x_{3j} \tau^y_{3j+1} + \tau^y_{3j+1} \tau^x_{3j+2} + \tau^x_{3j+2} \tau^x_{3j+3}  + 
\nonumber \\
& \tau^y_{3j} \tau^z_{3j+1} \tau^y_{3j+2} +  \tau^x_{3j+1} \tau^z_{3j+2} \tau^y_{3j+3} + \tau^y_{3j+2} \tau^z_{3j+3} \tau^x_{3j+4} \bigr) \ .
\end{align}
where the $\tau$'s are the Pauli matrices. Terms of this type can be viewed as generalized Dzyaloshinskii-Moriya interactions,\cite{Dzyaloshinskii58,Moriya60} and they could be generated, for instance, with cold atoms in optical lattices\cite{Pachos04}.

The starting point of our construction of the hierarchy of $so(N)_1$ models is a system of $N$ decoupled critical TFI chains. The criticality of the latter is described by a product of $N$ Ising CFTs, that is related to the $so(N)_1$ CFT via a condensation transition. Drawing on this insight,\cite{Mansson13} the first step is to couple the TFI chains together in a non-local manner that respects all the symmetries of the system. This is in agreement with the observation by Witten\cite{Witten84} that $so(N)_1$ models are essentially equivalent to
$N$ Ising models at criticality. The second step is to employ spin duality transformations to write the coupled systems in a translationally invariant and Jordan-Wigner solvable form for arbitrary $N$. The models we obtain are fine-tuned to criticality by construction. By introducing generic couplings, we find that for certain parametrizations, for instance varying the relative couplings between the 2- and 3-spin terms in \rf{H3_first}, their phase diagrams can be viewed as generalizations of either of the two simplest models in our hierarchy, the transverse field Ising or the XY model. 

While the exact solvability is an attractive feature of our hierarchy, we also show that our condensation transition motivated approach is not limited to exactly solvable models. To this end we consider the one-dimensional quantum version of the the two-dimensional classical
Blume-Capel model\cite{blume66,capel66}, which roughly speaking is a spin-1 generalization of the TFI chain in a generic magnetic field. This model admits no known solution, but it has been shown to exhibit a tri-critical point described by the tri-critical Ising CFT. Applying the same construction as we did in the case of the $so(N)_1$ hierarchy, we derive another model with a critical point described by the predicted supersymmetric minimal model. 

We have structured the paper such that it is accessible for readers with different backgrounds. We start in Sec.~\ref{sec:method} with a general
description of the motivation underlying our construction, namely that of the condensate-induced transitions, and give an outline how it is applied to construct the hierarchy of $so(N)_1$ critical spin chains. This section gives the readers who are mainly interested in the resulting spin chains an idea of the method, without having to go through the details. These details are given in Sec.~\ref{sec:framework}, where
we review the concepts of anyon models and condensate-induced transitions between topological phases. We illustrate these concepts with examples that
are relevant for the current paper. In Sec.~\ref{sec:counterpart}, we apply these concepts to construct a general Hamiltonian counterpart of a condensation transitions in the setting of $N$ decoupled critical TFI chains. This hierarchical construction is the first main result of our work. The second main result is presented in Sec.~\ref{sec:hierarchy}, where we give the spin duality transformations to cast the hierarchy of resulting Hamiltonians into an exactly solvable and translationally invariant form. The phase diagrams of the constructed spin chains are studied in Sec.~\ref{sec:phase-diagram}, where we show that they exhibit structure that is qualitatively similar to those of the transverse field Ising and XY chains. The generality of our condensation transition motivated construction is demonstrated in Sec.~\ref{sec:blume-capel}, where we provide an example of it being applied to a spin-1 model that does not  admit exact solution. We conclude with Section~\ref{sec:conclusions}, where the physical realization of our models, their relation to known problems and various interesting future directions are discussed.

For the sake of clarity, some of the details of our work are left for the appendices. In
Appendix~\ref{app:boundary-hamiltonians} we explicitly derive the generalized
boundary term that we use to derive the hierarchy of $so(N)_1$ spin chains. The general spin duality transformations we
use to cast the spin chains in a translationally invariant form are given in
Appendix~\ref{app:general-transform}. In Appendix~\ref{sec:cfttospectrum} we explain at a general level the connection between CFTs and the spectra of critical 1D models. Finally, Appendix~\ref{sec:characters} contains the details of the spectra predicted by $so(N)_1$
CFTs that describe the criticality of the hierarchy of spin chains we construct. 

\section{Background and overview of the method}
\label{sec:method}

In this section, we briefly explain the main idea behind our construction of exactly solvable spin chains with $so(N)_1$ critical
points. Our method is based on our previous work\cite{Mansson13} where we argued that condensate-induced transitions in two-dimensional topological phases\cite{Bais09} have a precise counterpart for critical spin chains. We showed using two specific examples that if the CFTs describing the two critical spin chains are related via the condensation mechanism, then, up to a spin duality transformation, the two spin chains differed only by a non-local term. We argued that this \emph{condensing boundary term} implemented a counterpart of condensate-induced transition in critical spin chains by constraining the boundary conditions in a specific manner. In particular, we argued that the constraints are equal to removing those states from the spectrum that corresponded to the CFT fields that are confined in the condensation framework following the condensation of a bosonic field. 

Here we fully exploit this insight to construct local, translationally invariant and Jordan-Wigner solvable spin chains with $so(N)_1$ critical
points. Our main example in Ref.~\onlinecite{Mansson13} was to show that in the presence of suitable condensing boundary term two decoupled critical transverse field Ising chains (TFI) could be exactly mapped to the critical XY chain, in agreement with the condensation framework relating the corresponding Ising and $u(1)_4$ CFTs. Our main result in the present work is to show that this is just one example of a larger hierarchy. Instead of just two TFI chains, we will start from $N$ decoupled TFI chains described by a Hamiltonian $H^N_{\rm TFI}$. Using the framework of condensate-induced transitions we then construct a generalized condensing boundary term $H^N_B$. This term couples all the TFI chains together in a non-local manner that constrains the boundary conditions such that the desired CFT fields are effectively confined. At the microscopic level this is equivalent to forcing all the $N$ TFI chains to have simultaneously either periodic or anti-periodic boundary conditions.

 The simple structure of this term enables us to find general duality transformations that allow us to write the total Hamiltonian 
 $H_{so(N)_1}=H^N_{\rm TFI}+H^N_B$ as a translationally invariant spin chain for arbitrary $N$. These spin chains are $N$-local 
 in the sense that the resulting Hamiltonians will always contain $N$-spin operators acting on up to $N$ adjacent spins. However, 
 the form of these many-body operators is such that all the constructed $H_{so(N)_1}$ models can always solved by means of a Jordan-Wigner 
 transformation. In terms of fermions the $N$-locality means that the unit cell grows linearly with $N$, with the fermions subject to 
 tunneling and paring of ranges up to $N-1$ nearest neighbors. By comparing the analytic solutions to the CFT predicted by the condensation framework, 
 we explicitly verify that the spin chains are critical and indeed described by the $so(N)_1$ CFT. Finally, we analyze the phase diagrams 
 of the $H_{so(N)_1}$  spin chains when their couplings are tuned away from criticality. We find that for a suitable parametrization, for all odd $N$ models the phase diagrams are qualitatively similar to that of a TFI chain, while for even $N$ they are similar to that of the XY chain. For generic couplings both series of models exhibit phases beyond these two simplest members of the hierarchy.

The exact solvability of the constructed spin chains follows from two properties. First, the elementary building blocks of our construction, 
the critical TFI chains, are exactly solvable. Second, the condensing boundary term $H^N_B$ respects all the symmetries of the TFI system and,
 while being manifestly non-local, acts locally on every symmetry sector of the system. However, the exact solvability is not required by 
 the condensation framework. To demonstrate that it applies also to spin chains which do not admit a solution via a Jordan-Wigner 
 transformation (or any exact solution to our knowledge), we consider a similar construction as above for spin-1 chains. Instead of TFI chains, we start from a decoupled system of two spin-1 Blume-Capel models described by $H_{\rm BC}$. Roughly speaking, these are spin-1 generalizations of a TFI chain in a general magnetic field and they are known to have a tri-critical point described by the tri-critical Ising CFT. We employ again the condensation framework to construct an appropriate condensing boundary term $H_B^{\rm BC}$. When the Blume-Capel chains are coupled by such a term and the $H_{\rm BC}+H^{\rm BC}_B$ system is fine-tuned to the tri-critical point, we verify that the critical behavior coincides with the predicted supersymmetric minimal model with central charge $c=7/5$. By constructing a duality transformation for the spin-1 system, we show that the Hamiltonian $H_{\rm BC}+H^{\rm BC}_B$ can be transformed into a local and translationally invariant form that bears striking similarity to the constructed hierarchy of spin-1/2 chains.

\section{The framework of condensate-induced transitions}
\label{sec:framework}

In this section we first introduce the minimal knowledge of anyon models and CFTs that is required to understand the framework of 
condensate-induced transitions. Then we illustrate the transitions using examples that motivate us later to construct the family of 
exactly solvable spin chains with $so(N)_1$ critical points. Readers interested only in the spin chains themselves can skip this section and
go directly to Section \ref{sec:counterpart}.

\subsection{Anyon models and CFTs of type $so(N)_1$}
\label{sec:anyonmodels}

The hallmark of two-dimensional topologically ordered phases is that their low-energy theories, regardless of the microscopics, are fully 
described by  topological quantum field theories, or more informally, by `anyon models'. For our purposes we can regard them as sets 
of data that encode the properties of the different types of quasiparticle excitations of the system. In a nutshell, an anyon model is specified by (i) the types of anyons (topological charges carried by the quasiparticles), (ii) their fusion rules (how can two quasiparticles behave when combined) and (iii) their topological spins (that encode the mutual statistics of the quasiparticles). A thorough account of such models can be found in Ref.~\onlinecite{k06}, but for our purposes only the minimal
information consisting of the three ingredients above is needed.

There is an intimate connection between anyon models and CFTs. This connection, usually going under the name of bulk-edge correspondence, 
states for a large class of models that if the gapped bulk of a two-dimensional topologically ordered phase is described by a given anyon model, then the gapless
 one-dimensional edge of the system is described by a given CFT, and vice versa.  The precise correspondence between the data characterizing 
 the anyon model and the CFT is as follows: The different anyon types correspond to the primary fields of the CFT, which both satisfying 
 the same sets of fusion rules (in the latter they appear as expansions of primary field correlation functions). Each field $a$ is also 
 associated with scaling dimension $h_a$ that is directly related to the topological spin of the anyon $a$ through $\theta_a = e^{2 \pi i h_a}$. 
 For a comprehensive account of CFTs, we refer to Ref.~\onlinecite{byb}. An additional important property of a given CFT is its central charge $c$, which is a measure of the number of degrees of freedom in the theory. 

We will illustrate these concepts below with few examples that are relevant to us. In particular, we will be mainly concerned with CFTs which go under the name of $so(N)_1$. These models have $N$ real fermionic degrees of freedom, corresponding to a central charge $c=N/2$, where each
real fermion contributes $c=1/2$ (a single bosonic degree of freedom would contribute $c=1$). Their primary fields, fusion rules and scaling dimensions exhibit systematic structure that we consider separately for $N$ odd and $N$ even.

\subsubsection{$so(N)_1$ CFTs with $N$ odd}

All $so(N_1)$ CFTs with odd $N$ contain three primary fields that we denote by $1$, $\psi$ and $\sigma$.
Their scaling dimensions are $h_1 = 0$, $h_\psi=1/2$ and $h_\sigma=N/16$, respectively,
and they satisfy the fusion rules (which are commutative, $a \times b = b \times a$)
\bq \label{fusion_Nodd}
  \psi \times \psi = 1, \quad \sigma \times \psi = \sigma, \quad \sigma \times \sigma = 1 + \psi,
\eq
with $\psi$ and $\sigma$ fusing trivially with the vacuum $1$, i.e. $\psi \times 1 = \psi$ and $\sigma \times 1 = \sigma$.
From these fusion rules and scaling dimensions we infer that as an anyon model $1$ would denote the vacuum state and
$\psi$ a fermionic quasiparticle (half-integer spin). On the other hand, $\sigma$ would correspond to a non-Abelian anyon, because of the fractional spin and because fusing $\sigma$ with itself has several possible outcomes. 

Due to isomorphisms between the associated CFTs, some of the theories go under more
familiar names. For instance, $so(1)_1$ is more commonly known as the Ising CFT, because
it is the CFT describing the critical point of the two-dimensional classical Ising model
(as well as the critical point of the one-dimensional transverse field Ising chain). As an anyon model it is relevant, for instance, to $p$-wave superconductors where $\psi$ corresponds to a Bogoliubov quasiparticle and $\sigma$ to a Majorana mode binding vortex.\cite{Read00}
In addition, $so(3)_1$ is usually referred to as $su(2)_2$.

\subsubsection{$so(N)_1$ CFTs with $N$ even}

All $so(N)_1$ CFTs with even $N$ contain four primary fields, $1$, $\psi$, $\lambda_1$ and $\lambda_2$, with scaling dimensions
$h_1 = 0$, $h_\psi=1/2$ and $h_{\lambda_1}=h_{\lambda_2}=N/16$, respectively. The fusion rules depend on $N$ such that for $N=2,6,\ldots$ they are given by
\bq \label{fusion_Neven1}
\psi \times \psi = 1, \quad
\lambda_1 \times \psi = \lambda_2, \quad
\lambda_2 \times \psi = \lambda_1,
\nonumber \\
\lambda_1 \times \lambda_1 = \lambda_2 \times \lambda_2 = \psi, \quad
\lambda_1 \times \lambda_2 = 1, 
\eq
while for $N=4,8,\ldots$ they are given by
\bq \label{fusion_Neven2}
\psi \times \psi = 1, \quad
\lambda_1 \times \psi = \lambda_2, \quad
\lambda_2 \times \psi = \lambda_1,
\nonumber \\
\lambda_1 \times \lambda_1 = \lambda_2 \times \lambda_2 = 1, \quad
\lambda_1 \times \lambda_2 = \psi.
\eq
As an anyon model, $\psi$ would again be a fermion, whereas $\lambda_1$ and $\lambda_2$ would be Abelian anyons except for $N$ an odd
multiple of $8$, in which case they are fermions, and for $N$ a multiple of $16$, in which case they are bosons.

Like the odd $N$ cases, some of the even cases are known more commonly under other names.
The $so(2)_1$ CFT, describing the criticality of the XY spin chain, is often denoted as $u(1)_4$, while 
$so(4)_1$ can be denoted as a product theory  $u(1)_2 \times u(1)_2 \simeq su(2)_1 \times su(2)_1$. As anyon models both are relevant, for instance, to collective vortex states in $p$-wave superconductors.\cite{Lahtinen12}

\subsubsection{Product theories}

In this paper, we make frequent use of products of CFTs that can be constructed in a straightforward manner. For instance,
a direct product theory of $N$ Ising CFTs with central charge $c = N/2$, denoted here by Ising$^{\times N}$ $\equiv$ Ising $\times \cdots \times$ Ising, consists of $3^N$ primary fields that are labeled as $(a_1,\ldots,a_N)$, with $a_1,\ldots,a_N = 1,\psi$ or $\sigma$. The fusion rules
of these fields follow associatively from the fusion rules of a single Ising CFT and the scaling dimensions are obtained as the sum of those of the constituent fields, i.e. $h_{(a_1,\ldots,a_N)} = \sum_{i=1}^N h_{a_i}$.

\subsection{Condensate-induced transitions between topological phases}
\label{sec:cond-ex}

A condensate-induced transition occurs when a bosonic quasiparticle in a topologically ordered phase condenses. Without going into the microscopic details of such a process, the nature of the \emph{condensed phase} can be worked out at the level of anyons models, as has been studied in detail in Ref.~\onlinecite{Bais09}. Condensation implies that the vacuum state is redefined, which imposes consistency conditions on the other quasiparticles in the system. These conditions derive from demanding that the condensate of the bosonic quasiparticles behaves like a genuine vacuum, i.e. that
\begin{itemize}
\item[(i)]
It fuses trivially with all other quasiparticles
\item[(ii)]
It has trivial statistics with all other quasiparticles
\item[(iii)]
It is unique.
\end{itemize}
Violating any of these conditions means that the quasiparticle spectrum must change (some particles are identified, some confined)
in a manner that leads to these three conditions being satisfied.\cite{Bais09} We illustrate the condensate-induced transitions with three examples that are relevant to us.

\subsubsection{${\rm Ising}^{\times 2}  \to so(2)_1$} 

As the first example, we consider the condensation of the boson $(\psi,\psi)$  in the Ising$^{\times 2}$ CFT.
The first step of condensation is to identify this boson with the vacuum label, i.e. we set $(\psi,\psi) = (1,1)$.
In order for $(\psi,\psi)$ to behave like the vacuum, the demand (i) above implies that all particles $a$ and $b$ that are related by fusion with the boson should be identified. That is, if $a \times (\psi,\psi) = b$, then we set $a = b$. We arrive at a reduced set of particle types $\tilde{1} = (1,1) = (\psi,\psi)$,
$\tilde{\psi}=(1,\psi)=(\psi,1)$, $\tilde{\sigma}_1=(\sigma,1)=(\sigma,\psi)$ and $\tilde{\sigma}_2=(1,\sigma)=(\psi,\sigma)$,
while the particle $(\sigma,\sigma)$ remains unaffected at this step.

Demanding (ii) is equivalent to confining all particles that have non-trivial statistics with the new vacuum. This in turn is equivalent to removing all identified particles with unequal conformal weights from the spectrum\cite{Bais09}. Since $h_{(\sigma,1)}=1/16$, but $h_{(\sigma,\psi)}=9/16$, the particles $\tilde{\sigma}_1$ and $\tilde{\sigma}_2$ are eliminated from the particle content of the condensed phase.

Finally, demanding (iii) one finds that $(\sigma,\sigma)$ has to branch into several particles, because fusion with itself gives rise to two times the new vacuum,
$
(\sigma,\sigma) \times (\sigma,\sigma) = (1,1) + (1,\psi) + (\psi,1) + (\psi,\psi) = 2 \cdot \tilde{1} + 2 \cdot \tilde{\psi}.
$
The uniqueness of the vacuum can be satisfied if one replaces $(\sigma,\sigma)$ by $\lambda_1  + \lambda_2$ and demands that the particles
 $\tilde{1}$, $\tilde{\psi}$, $\lambda_1 $ and $\lambda_2$ satisfy the fusion rules \rf{fusion_Neven1}. Evaluating the scaling dimensions of
  these particles as sums of the constituent ones, we obtain $so(2)_1 \simeq u(1)_4$ as the self-consistent theory for the condensed phase.

\subsubsection{$so(2)_1 \times {\rm Ising}  \to so(3)_1$} 

As the second example we consider the condensation of the boson in the
$so(2)_1 \times$ Ising theory. We label the particles in the $so(2)_1$ theory as in the previous example by
$\tilde{1},\lambda_1,\lambda_2,\tilde{\psi}$. The product theory $so(2)_1 \times {\rm Ising}$
contains a boson $(\tilde{\psi},\psi)$, which we identify with the vacuum $(\tilde{1},1)$.
Going through the steps (i)-(iii) again as described above, we find
first that the following particles are pairwise  identified:
$(\tilde{1},1) = (\tilde{\psi},\psi)$,
$(\tilde{1},\psi) = (\tilde{\psi},1)$,
$(\lambda_1,\sigma) = (\lambda_2,\sigma)$
and
$(\tilde{1},\sigma) = (\tilde{\psi},\sigma)$,
$(\lambda_1, 1) = (\lambda_2,\psi)$,
$(\lambda_1, 1) = (\lambda_1,\psi)$.
However, the identified particles in the latter set have unequal scaling dimensions, which means that they have to be confined.
The fusion rules of the remaining three particles are self-consistent and the new vacuum is unique. Thus, no particles need to split and one obtains a theory with three particles: $1' = (\tilde{1},1) = (\tilde{\psi},\psi)$, $\psi' = (\tilde{1},\psi) = (\tilde{\psi},1)$ and $\sigma' = (\lambda_1,\sigma) = (\lambda_2,\sigma)$. These particles satisfy the fusion rules in Eq.~\eqref{fusion_Nodd}, and by
considering the scaling dimensions, one finds that the condensed phase is described by the $so(3)_1 \simeq su(2)_2$ model.

\subsubsection{Generalization: ${\rm Ising}^{\times N} \to so(N)_1$}

Since the main difference between different $N$ even and $N$ odd theories lies in the different $N$ dependent scaling dimensions, we can immediately see that the two examples above obey the general rule: If one condenses the boson formed out of the two fermions in a $so(N)_1 \times so(N')_1$ theory, then one induces the transition
\be
  so(N)_1 \times so(N')_1 \to so(N+N')_1.
\ee
Recalling that Ising$^{\times 3} \simeq so(1)_1^{\times 3}$, this implies that the two examples above can be viewed as two successive condensations 
\be
  so(1)_1^{\times 3} \to so(2)_1 \times so(1)_1 \to so(3)_1.
\ee 
It is then straightforward to generalize the result of successive condensations in the $so(1)_1^{\times N}$ model for arbitrary $N$. 
Without loss of generality, we can always choose to condense consecutively and pairwise $N-1$ bosons such that the process is equivalent 
to a series of alternating pairwise condensations between product theories consisting of either of  two odd $N$ theories (Example 1) or 
an even $N$ and an odd $N$ theory (Example 2). Regardless of the order in which the bosons are condensed, the fully condensed phase will be described by the $so(N)_1$ theory. At the level of CFTs this means that condensate-induced transitions conserve the central charge, i.e. both the uncondensed and condensed phases have the same number of degrees of freedom.

Our main result is to show that this same structure applies also to critical spin chains. Motivated by the condensation framework, we will start with $N$ decoupled critical TFI chains (described by Ising$^{\times N}$) and implement the counterpart of condensation by coupling them with a non-local condensing boundary term. Employing a suitable duality transformation we arrive at a local spin chain with an $so(N)_1$ critical point.

\section{Condensate-induced transitions and critical spin chains}
\label{sec:counterpart}

In this section we first introduce the elementary building block of our construction -- the critical transverse field Ising chain. Then we review the results of Ref.~\onlinecite{Mansson13} that provide the physical motivation for the construction of the condensing boundary term. Going beyond the earlier studies for two critical TFI chains, we generalize the condensing boundary term for a system of $N$ decoupled TFI chains.

\subsection{The critical transverse field Ising chain}
\label{sec:critical-tfi}

The starting point of our derivation of a spin chain, with an
$so(N)_1$ critical point, is a system of $N$ decoupled critical TFI chains. We realize such a system as a single critical TFI chain of 
length $L$ (with $L$ a multiple of $N$) with $N^{\rm th}$ nearest neighbor interactions only
\be \label{NTFI}
  H^N_{TFI} = \sum_{j=0}^{L-1} \sigma^x_{j} \sigma^x_{j+N} + \sigma^z_{j}.
\ee
Periodic boundary conditions, i.e. $\sigma^{\alpha}_{j+L} = \sigma^{\alpha}_{j}$, are always assumed. For $N=1$ one recovers the usual 
nearest neighbor TFI chain, which is readily solved via a Jordan-Wigner transformation \cite{epw70,p70}. The resulting fermionic 
Hamiltonian is given by
\begin{align} \label{1TFI_JW}
H^{1}_{\rm TFI} =&
\sum_{j=0}^{L-1} (2c^{\dagger}_{j} c^{\vphantom{\dagger}}_{j}-1) +
\sum_{j=0}^{L-2}
(c^{\vphantom{\dagger}}_{j} - c^{\dagger}_{j}) (c^{\vphantom{\dagger}}_{j+1} + c^{\dagger}_{j+1})
\nonumber \\
&-\mathcal{P} (c^{\vphantom{\dagger}}_{L-1} - c^{\dagger}_{L-1})
(c^{\vphantom{\dagger}}_{0} + c^{\dagger}_{0})\, ,
\end{align}
where $c_j$ is a complex fermion operator at site $j$. This Hamiltonian describes paired fermions and conserves only the fermion parity 
described by the symmetry operator
$\mathcal{P} = \prod_{j=0}^{L-1} \sigma^{z}_i = \exp(i\pi \sum_j c_j^\dagger c_j)$.
The presence of the operator $-\mathcal{P}$, multiplying the hopping and interacting term
crossing the boundary, leads to a coupling between the parity sectors and the boundary
conditions.
For odd parity ($\mathcal{P}=-1$) one has periodic boundary conditions ($c_{L} \equiv c_{0}$), while for even parity ($\mathcal{P}=1$), one has
 anti-periodic boundary conditions  ($c_{L} \equiv -c_{0}$). In momentum space the Hamiltonian can be diagonalized with a Bogoliubov 
 transformation, which gives the spectrum 
\begin{align} 
H_{\rm TFI}^1 &=
\sum_{k}
\sqrt{2 + 2 \cos \bigl( 2 \pi k /L \bigr) }
\bigl(2 c^{\dagger}_{k} c^{\vphantom{\dagger}}_{k} - 1 \bigr)  \ , 
\label{1TFI_E}
\end{align}
where the $c^{\dagger}_{k}$ create fermions with momentum $k$\cite{footnote1}.
Due to the parity-dependent boundary conditions, these momenta take
integer values for $\mathcal{P}=-1$ and half-integer values for $\mathcal{P}=1$.

The CFT describing the criticality of the TFI chain is the Ising CFT with central charge $c=1/2$. This CFT has three primary fields, $1$, 
$\sigma$ and $\psi$, with scaling dimensions $h_1 = 0$, $h_\sigma = 1/16$ and $h_\psi = 1/2$.
In general, if the spectrum of a critical quantum chain can be described in terms of a CFT, this
implies that all the states can be labeled by the fields (or sectors) of the CFT.
The energy of these states (after an appropriate shift and rescaling) take the form $E = 2 h + n$ in the large $L$ limit.
Here, $h$ denotes the scaling dimension of a primary field in the CFT, and $n$ is a non-negative
integer. The connection between CFT and critical spectra is discussed in more detail in the Appendix~\ref{sec:cfttospectrum}.
For our purposes the essential property is the correspondence between the assignment of the
CFT sectors and the boundary conditions. 
In particular, all states in the even parity sector ($\mathcal{P} = 1$, with anti-periodic boundary conditions) are labeled by 
either $1$ or $\psi$, while all the states the odd parity sector ($\mathcal{P} = -1$, with periodic boundary conditions) are labeled by $\sigma$.

For $N>1$ the Hamiltonian \rf{NTFI} describes $N$ completely decoupled critical TFI chains of length $L/N$. Each of these chains can 
independently be solved via a Jordan-Wigner transformation.  The parity of the fermions is conserved independently for each of the $N$ chains,
which means that $H^N_{TFI}$ has $N$ mutually commuting symmetry operators given by
\be
\label{eq:Pparity}
  \mathcal{P}_n = \prod_{j=0}^{L/N-1} \sigma^z_{jN+n}, \qquad n=0,1,\ldots,N-1.
\ee
The boundary conditions for the fermions in chain $n$ depend only on the parity $\mathcal{P}_n$ of the fermions in that chain.
When diagonalized, the dispersion relation for each chain will be identical to \rf{1TFI_E}. Since the $N$ TFI chains are decoupled, the criticality of 
the whole system is described by the Ising$^{\times N}$ CFT with central charge $c=N/2$. The labeling of the states by the Ising$^{\times N}$ 
primary fields follows directly from the labeling of the states in each chain according to the correspondence between the boundary conditions 
described above.

\subsection{The condensing boundary term}
\label{sec:boundaryterm}

We argued in Ref.~\onlinecite{Mansson13} that the counterpart of condensate-induced transitions in critical spin chains occurs not through 
the condensation per se, but through the confinement of some of primary fields it induces. Because of the correspondence between the CFT 
sectors and the boundary conditions of the TFI chains, constraining the set of allowed boundary conditions is equivalent to removing some 
of the CFT sectors from the theory. Our main result in Ref.~\onlinecite{Mansson13} was to argue that in 
a system of two critical TFI chains the states labeled by the confined CFT primary fields could in general be removed from the spectrum by 
adding to the Hamiltonian a non-local term that made the boundary conditions of the two chains depend also on the symmetry sectors of each 
other. We showed that the spectrum is subsequently modified in a manner that encoded all the features of condensate-induced transitions and 
had the predicted critical behavior.

\begin{table*}[th]
\begin{tabular}{c}
Sectors of $H^2_{\rm TFI}$\\
\begin{tabular}{|c|c|c|}
\hline
 $(\mathcal{P}_0,\mathcal{P}_1)$ & $ (BC_0,BC_1)$  & ${\rm Ising}^{\times 2}$ fields \\
	\hline
(1,1) & (-1,-1) & $(1,1)$, $(1,\psi)$, $(\psi,1)$, $(\psi,\psi)$ \\
\hline
(1,-1)  & (1,-1)  & $(1,\sigma)$, $(\psi,\sigma)$ \\
\hline
(-1,1) & (-1,1) & $(\sigma,1)$, $(\sigma,\psi)$ \\
\hline
(-1,-1)  & (1,1)  & $(\sigma,\sigma)$ \\
\hline
\end{tabular}
\end{tabular}
\begin{tabular}{c}
Sectors of $H^2_{\rm TFI}+H^2_B=H_{XY}$\\
\begin{tabular}{|c|c|c|c|c|}
\hline
 $(\mathcal{P}_0,\mathcal{P}_1)$ & $ (BC_0,BC_1)$  & ${\rm Ising}^{\times 2}$ fields & $\mathcal{T}^z$ & $u(1)_4$ fields \\
	\hline
(1,1) & (-1,-1) & $(1,1)$, $(1,\psi)$, $(\psi,1)$, $(\psi,\psi)$ & 1 & $1,\tilde{\psi}$ \\
\hline
(1,-1)  & (1,1)  & $(\sigma,\sigma)$ & -1 & $\lambda,\bar{\lambda}$\\
\hline
(-1,1) & (1,1) & $(\sigma,\sigma)$ & -1 & $\lambda,\bar{\lambda}$ \\
\hline
(-1,-1)  & (-1,-1)  & $(1,1)$, $(1,\psi)$, $(\psi,1)$, $(\psi,\psi)$ & 1 & $1,\tilde{\psi}$ \\
\hline
\end{tabular}
\end{tabular}
\caption{\label{Cond2}%
{\it Left:} The symmetry sectors $(\mathcal{P}_0,\mathcal{P}_1)$, the corresponding boundary conditions $(BC_0,BC_1)$ in the fermionic picture and the Ising$^{\times 2}$ CFT sectors that label all the states in the respective symmetry sectors of $H^2_{TFI}$.
{\it Right:} Same, but now in the presence of the confining boundary term $H^2_B$. After performing the spin duality transformations \rf{Transform2} to map the system to \rf{H2},  the symmetry sectors mix and map to the parity symmetry sectors of the XY chain described by the operator $\mathcal{T}^z=\mathcal{P}_0\mathcal{P}_1$. In Ref.~\onlinecite{Mansson13} we showed that the matching of the states in the two models is in exact agreement with the predictions of the condensation framework.}
\end{table*}

To motivate the generalization to a system of $N$ decoupled TFI chains, let us briefly revisit this example where the condensing boundary 
term takes the form
\be \label{HB2}
 H_B^2 = \bigl( \mathcal{P}_1 - \id \bigr) \sigma^{x}_{L-2}\sigma^{x}_{0} +
 \bigl( \mathcal{P}_0 - \id \bigr) \sigma^{x}_{L-1}\sigma^{x}_{1}\,.
\ee
Adding it to $H^2_{\rm TFI}$ we obtain  
\begin{align}
 H^2_{\rm TFI} + H_B^2 = &
\Bigl(
\sum_{j=0}^{L-3} \sigma^x_{j} \sigma^x_{j+2} + \sigma^z_{j} \Bigr) + \\
&\mathcal{P}_1 \sigma^{x}_{L-2}\sigma^{x}_{0} + \mathcal{P}_0 \sigma^{x}_{L-1}\sigma^{x}_{1} \ . \nonumber
\end{align}
If we would solve this problem using a Jordan-Wigner transformation, the boundary conditions for the fermions in both chains would now 
depend on the product $\mathcal{P}_0 \mathcal{P}_1$. This means instead of four independent boundary conditions, both chains are now forced 
to have simultaneously either periodic or anti-periodic boundary conditions. As illustrated in Table~\ref{Cond2},  the correspondence between 
the boundary conditions and the CFT sectors is modified such that there are no longer states labeled by confined primary fields in the 
spectrum (see Section~\ref{sec:cond-ex} for the corresponding condensate-induced transition). Indeed, by studying how the energy spectrum 
is precisely modified (states in some symmetry sectors occur now at integer momenta while they used to occur for half-integer momenta, and 
vice versa), we found that it coincides precisely with the critical spectrum described by the $so(2)_1 \simeq u(1)_4$ CFT. Moreover, 
we showed by an exact mapping that $H^2_{\rm TFI} + H_B^2$ is equivalent to a critical XY chain\cite{Mansson13}
(see also Sec.~\ref{sec:caseN2}).

The key insight behind the form of the condensing boundary term was the correspondence between the boundary conditions and the CFT 
sectors. This correspondence applies beyond the TFI chain. For instance, for the critical XY chain described by the $so(2)_1$ CFT the 
primary fields $1$ and $\psi$ again always label states for anti-periodic boundary conditions, while all the states for periodic boundary 
conditions are labeled by either $\lambda_1$ or $\lambda_2$. As we discussed in Section~\ref{sec:cond-ex}, condensate-induced transition 
in an Ising$^{\times N}$ system can always be understood as a series of consecutive pairwise transitions between two odd $N$ theories 
(like the Ising$^{\times 2}$ example above) or between an $N$ even and $N$ odd theories. The simplest example of the latter in spin chains is 
realized in a decoupled system of a critical TFI chain and a critical XY chain that is described by Ising$\times so(2)_1$ CFT. In 
Appendix~\ref{app:boundary-hamiltonians} we show that this transition can be realized by coupling the two chains with a condensing boundary 
term similar in form to \rf{HB2}. Since the XY chain can be viewed as emerging from two coupled TFI chains as described above, we find that we could 
have equally started with a system of three decoupled critical TFI chains and added the condensing boundary term 
\bq \label{HB3}
 H^3_{B} & = & (\mathcal{P}_1 \mathcal{P}_2-\id)\sigma^x_{L-3}\sigma^x_0 + (\mathcal{P}_0 \mathcal{P}_2-\id) \sigma^x_{L-2}\sigma^x_1 + \nonumber \\ 
   \ & \ & + (\mathcal{P}_0\mathcal{P}_1-\id)\sigma^x_{L-1}\sigma^x_2.
\eq
Like in the $N=2$ case, this term couples the TFI chains such that when $H^3_{\rm TFI}+H^3_B$ is fermionized with a Jordan-Wigner transformation, all three are forced again to have simultaneously either periodic or anti-periodic boundary conditions. 

This motivates us to write down a condensing boundary term that implements the counterpart of successive condensations of all the $N-1$ bosons in a system of $N$ decoupled TFI chains. Assuming that $L/(2N)$ is an integer (the $L$-dependent form is given in App.~\ref{app:boundary-hamiltonians}), the general condensing term is given by
\be \label{HBN}
	H_B^N = \sum_{n=0}^{N-1} \left[\left(\prod_{l \neq n}\mathcal{P}_l-\id \right) \sigma^x_{L-N+n} \sigma^x_{n}\right].
\ee
We have verified, for $N \leq 16$, that the Hamiltonians $H_{TFI}^N+H^N_B$ indeed are critical, and always have the $so(N)_1$ critical behavior 
predicted from the condensation picture, by making use of the exact solution we present in the next section.

The non-local Hamiltonians $H_{TFI}^N+H^N_B$ seem to break translational invariance. However, when restricted to any one of the symmetry 
sectors $(\mathcal{P}_0,\mathcal{P}_1,\ldots,\mathcal{P}_{N-1})$, they are local and translationally invariant. This suggests that it could be 
possible to find duality transformations that would give local and translationally invariant representation for all these Hamiltonians. Indeed, 
in the next section we provide such transformations for arbitrary $N$ and show that the resulting Hamiltonians can be solved by means of a 
Jordan-Wigner transformation.

\section{Exactly solvable spin chains with $so(N)_1$ critical points}
\label{sec:hierarchy}

Motivated by the framework of condensate-induced transitions, we argued above that spin chains of the form $H_{TFI}^N+H^N_B$ are always 
critical and described by $so(N)_1$ CFT. We now set the condensation picture aside and focus on constructing local, translationally 
invariant and Jordan-Wigner solvable representations for these spin chains. 

To this end we employ a duality transformation between 
the $\sigma$ spin variables that used to write down the TFI chains and a new set Pauli operators $\tau$. In particular, we will show that
\be \label{HN}
  H_{so(N)_1}(\tau) = H_{TFI}^N(\sigma)+H^N_B(\sigma),
\ee
where $H_{so(N)_1}(\tau)$ is $N$-local (some operators act on $N$ adjacent spins) and translationally invariant with respect to a unit cell of $N$ ($N$ odd) or $N/2$ sites ($N$ even). The form of the $N$-local operators is such that the spin Hamiltonians can be solved by the means of a Jordan-Wigner transformation. Like for the TFI chains, the boundary conditions for the fermions turn out always to depend on the total fermion parity, which turns out to take the general form
\be \label{PN}
\mathcal{P}_{so(N)_1} = \prod_{n=0}^{N-1} \mathcal{P}_n = \prod_{j=0}^{L-1} \sigma^z_{j} = \prod_{j=0}^{L-1} \tau^z_{j}.
\ee
We recall that the parity operators $\mathcal{P}_n$ are defined in Eq.~\eqref{eq:Pparity}.
In other words, the condensing boundary term couples the $N$ TFI chains in such a way that their joint fermion parity coincides always 
with the parity of the  $H_{so(N)_1}$ chain. The duality transformations required to transform the spin chains to a local form respect 
this in the sense that in terms of both $\sigma$ and $\tau$ operators the parity operator $\mathcal{P}_{so(N)_1}$ takes the same form. 
Another general thing to note is that since $H^N_B$ does not break any of the $\mathcal{P}_n$ symmetries, they must also be symmetries 
of $H_{so(N)_1}$. Thus the number of symmetry operators for $H_{so(N)_1}$ increases linearly with $N$, which results in an increased 
degeneracy in their spectra.

We will first present the explicit forms of the $H_{so(N)_1}$ Hamiltonians, the required duality transformations to obtain them and 
their solutions for $N \leq 5$. Based on their systematic form we then give the general form for exactly solvable $H_{so(N)_1}$ Hamiltonians 
and their solutions for arbitrary $N$. As the corresponding general duality transformations are lengthy, we present them in 
Appendix~\ref{app:general-transform}.

\subsection{The $N=2$ case}
\label{sec:caseN2}

The simplest case of $N=2$ has been considered in Ref.~\onlinecite{Mansson13},
where it is shown that the Hamiltonian $H_{so(2)_1}$ 
coincides exactly with that of a critical XY chain.
This follows from the spin duality transformations
\begin{align}
\label{Transform2}
\sigma^{z}_{2j} &= \tau^{y}_{2j} \tau^{y}_{2j+1} &
\sigma^{z}_{2j+1} &= \tau^{x}_{2j} \tau^{x}_{2j+1} \\
\nonumber
\sigma^{x}_{2j} &= (\prod_{i<j} \tau^{x}_{2i} \tau^{x}_{2i+1}) \tau^{x}_{2j} &
\sigma^{x}_{2j+1} &= \tau^{y}_{2j+1} (\prod_{i>j} \tau^{y}_{2i}\tau^{y}_{2i+1}) \, ,
\end{align}
that when applied to \rf{HN}, give the exact relation
\be \label{H2}
   H_{so(2)_1} = 
  \sum_{j=0}^{L-1} \left( \tau^x_{j} \tau^x_{j+1} + \tau^y_{j} \tau^y_{j+1} \right) .
\ee
We note that transformations similar to Eq.~\eqref{Transform2}, and the relation between two TFI chains and the XY chain for open boundary conditions, were studied in Refs.~\onlinecite{capel77,jf78,pjp82,f94}.

The solution after a Jordan-Wigner transformation is given by
\begin{equation}
H_{so(2)_1} = \sum_{k} \sqrt{2+ 2 \cos \Bigl( \frac{4 \pi k} {L} \Bigr)} \bigl( 2c^{\dagger}_{k} c^{\vphantom{\dagger}}_{k} -1 \bigr),
\end{equation}
where the momenta $k$ depend on the parity $\mathcal{P}_{so(2)_1}$ of fermions present in the system. For odd parity ($\mathcal{P}_{so(2)_1}=-1$), the momenta take integer values $k=0,1,\ldots, L-1$, while for even parity ($\mathcal{P}_{so(2)_1}=1$), the momenta take half-integer values $k=1/2, 3/2, \ldots, L-1/2$. 


\subsection{The $N=3$ case}

For the case $N=3$ we employ the duality transformations (here $0 \leq i,j \leq L/3-1$)
\begin{widetext}
\begin{align}
\label{Transform3}
\sigma^{z}_{3j} &= \tau^{y}_{3j} \tau^{z}_{3j+1} \tau^{y}_{3j+2} & \sigma^{x}_{3j} &= \left( \prod_{i<j} \tau^{y}_{3i} \tau^{x}_{3i+1} \tau^{z}_{3i+2} \right) \tau^{y}_{3j} \tau^{x}_{3j+1} \tau^{y}_{3j+2} \left( \prod_{i>j} \tau^{z}_{3i} \tau^{x}_{3i+1} \tau^{y}_{3i+2} \right) \nonumber \\ 
\sigma^{z}_{3j+1} &= \tau^{x}_{3j} \tau^{y}_{3j+1} & \sigma^{x}_{3j+1} &= \left( \prod_{i<j} \tau^{y}_{3i} \tau^{x}_{3i+1} \tau^{z}_{3i+2} \right) \tau^{y}_{3j} \\
\sigma^{z}_{3j+2} &= \tau^{y}_{3j+1} \tau^{x}_{3j+2} & \sigma^{x}_{3j+2} &= \tau^{y}_{3j+2} \left( \prod_{i>j} \tau^{z}_{3i} \tau^{x}_{3i+1} \tau^{y}_{3i+2} \right). \nonumber 
\end{align}
\end{widetext}
Applying them to \rf{HN} we obtain the Hamiltonian
\begin{align} \label{H3}
  H_{so(3)_1} = \sum_{j=0}^{L/3-1} \Bigl( & \tau^y_{3j} \tau^z_{3j+1} \tau^y_{3j+2} + \tau^x_{3j+2} \tau^x_{3j+3} +  \nonumber \\
\ &  \tau^x_{3j+1} \tau^z_{3j+2} \tau^y_{3j+3} + \tau^x_{3j} \tau^y_{3j+1}  +  \\
\  & \tau^y_{3j+2} \tau^z_{3j+3} \tau^x_{3j+4} + \tau^y_{3j+1} \tau^x_{3j+2}  \Bigr),  \nonumber
\end{align}
which is translationally invariant with respect to a unit cell of three sites. Spin chains with similar Dzyaloshinskii-Moriya like interaction terms have been considered in the literature (see e.g. Refs.~\onlinecite{Gottlieb99,Zvyagin06}). However, unlike those models, our hierarchy models exhibit a particular alternation in the form of the couplings that underlies their systematic critical behavior.

Before performing a Jordan-Wigner transformation, it is useful to perform an additional transformation such that the fermionic form of the Hamiltonian will have real coefficients. A unitary operator implementing this acts on the sites $3j+1$, by swapping $\tau^x_{3j+1} \leftrightarrow \tau^y_{3j+1}$ and adding a sign as $\tau^z_{3j+1} \rightarrow -\tau^z_{3j+1}$. This gives the Hamiltonian
\be \label{H3_2}
  \hat{H}_{so(3)_1}=\sum_{j=0}^{L-1} \left( \mathcal{S}_j\tau^y_{j} \tau^z_{j+1} \tau^y_{j+2} + \tau^x_{j} \tau^x_{j+1} \right),
\ee
where $\mathcal{S}_j=-1$ for $j=0\mod 3$ and $\mathcal{S}_j=1$ otherwise. The full translational invariance is thus broken by the staggered sign of the three-spin interaction. After the Jordan-Wigner transformation, we obtain the Hamiltonian
(compare with the TFI case, Eq.~\eqref{1TFI_JW}), which consist of uniform nearest-neighbor and sign staggered
next nearest-neighbor hopping and pairing terms
\begin{widetext}
\bq
\label{N3_JW}
H_{so(3)_1} &= &\Biggl( \sum_{j=0}^{L/3-2} 
+ (c^\dagger_{3j} + c_{3j})(c^\dagger_{3j+2} - c_{3j+2})
- (c^\dagger_{3j+1} + c_{3j+1})(c^\dagger_{3j+3} - c_{3j+3})
- (c^\dagger_{3j+2} + c_{3j+2})(c^\dagger_{3j+4} - c_{3j+4}) \nonumber \\ &&
+ (c^\dagger_{3j+2} - c_{3j+2})(c^\dagger_{3j+3} + c_{3j+3})
+ (c^\dagger_{3j+3} - c_{3j+3})(c^\dagger_{3j+4} + c_{3j+4})
+ (c^\dagger_{3j+4} - c_{3j+4})(c^\dagger_{3j+5} + c_{3j+5}) \Biggr) \nonumber \\ &&
+ (c^\dagger_{L-3} + c_{L-3})(c^\dagger_{L-1} - c_{L-1})
+ \mathcal{P}_{so(3)_1} (c^\dagger_{L-2} + c_{L-2})(c^\dagger_{0} - c_{0})
+ \mathcal{P}_{so(3)_1} (c^\dagger_{L-1} + c_{L-1})(c^\dagger_{1} - c_{1}) \nonumber \\ &&
- \mathcal{P}_{so(3)_1} (c^\dagger_{L-1} - c_{L-1})(c^\dagger_{0} + c_{0})
+ (c^\dagger_{0} - c_{0})(c^\dagger_{1} + c_{1})
+ (c^\dagger_{1} - c_{1})(c^\dagger_{2} + c_{2}) \ .
\eq
\end{widetext}
Given that we started the derivation from
three decoupled TFI chains, and essentially only changed the boundary conditions
in some of the sectors, one can expect that the spectrum of $H_{so(3)_1}$ would bears close resemblance to the
spectrum of three TFI chains. Indeed, by Fourier transforming with respect to the three site unit cell and diagonalizing 
the six by six Bloch matrix, we obtain a spectrum in terms of three fermions $c_{n,k}$, with $n = 0,1,2$. To be precise, 
the spectrum is given by
\bq \label{H3E}
H_{so(3)_1} & = & \sum_{k}
\epsilon_{0,k}
\bigl( 2c^{\dagger}_{0,k} c^{\vphantom{\dagger}}_{0,k} -1 \bigr) +\\ 
\ & \ &
\epsilon_{1,k}
\bigl( 2c^{\dagger}_{1,k} c^{\vphantom{\dagger}}_{1,k} -1 \bigr) + 
\epsilon_{2,k}
\bigl( 2c^{\dagger}_{2,k} c^{\vphantom{\dagger}}_{2,k} -1 \bigr) \ ,\nonumber
\eq
where
\begin{equation*}
\epsilon_{n,k} =
\begin{cases}
\sqrt{2 + 2 \cos \bigl( \frac{2\pi k}{(L/3)}\bigr)} & \qquad \text{for } n = 0 \\
\sqrt{2 - 2 \cos \bigl( \frac{2\pi k}{(L/3)}\bigr)} & \qquad \text{for } n=1,2 \ .
\end{cases}
\end{equation*}
The momenta $k$ depend again on the total parity of fermions such that for $\mathcal{P}_{so(3)_1}=-1$ the momenta take integer 
values $k=0,1,\ldots, L/3-1$, while for $\mathcal{P}_{so(3)_1}=1$ the momenta take half-integer values $k=1/2, 3/2, \ldots, L/3-1/2$.

\begin{figure}[t]
\includegraphics[width=\columnwidth]{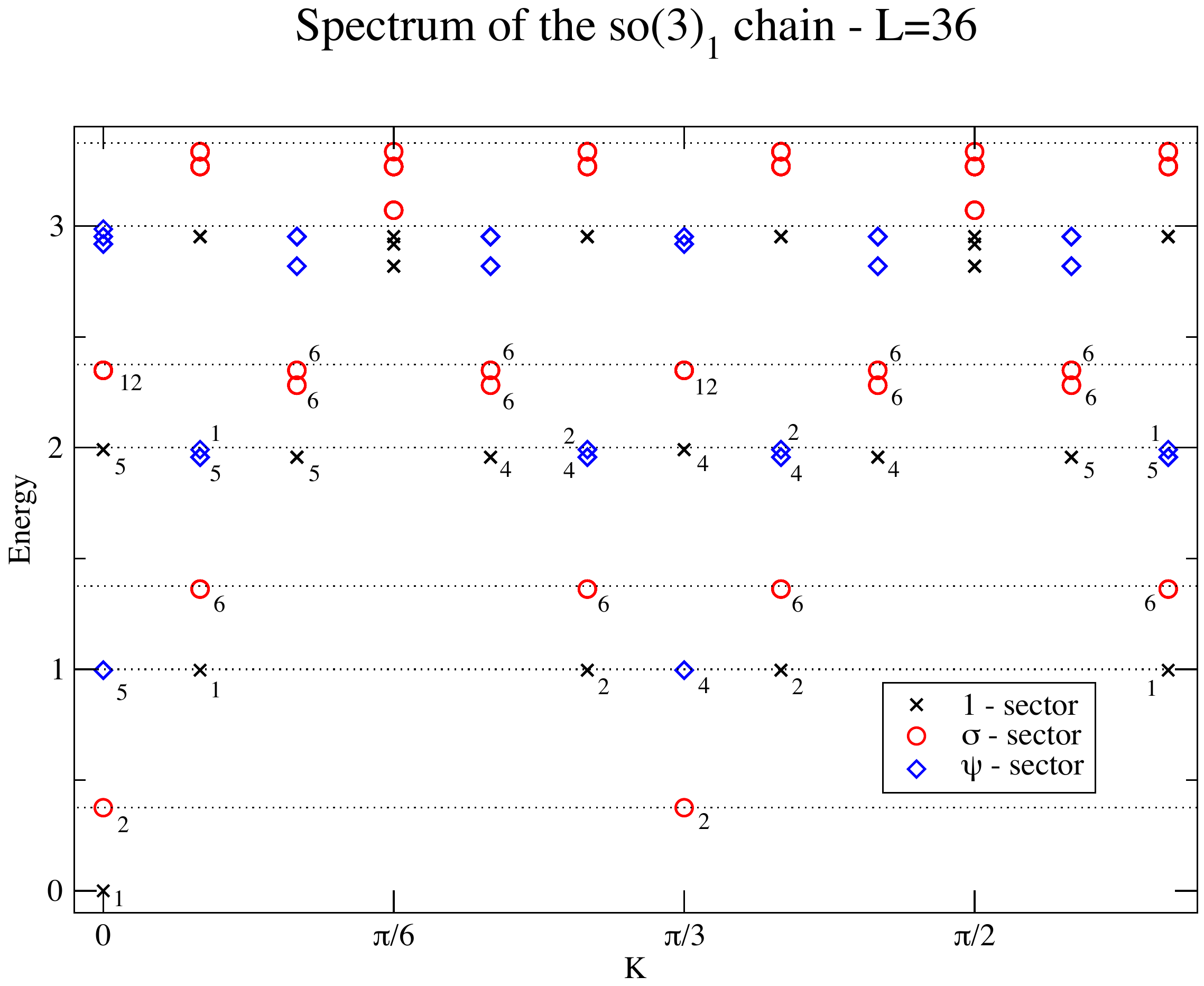}
\caption{The low-lying part of the rescaled spectrum of the critical spin chain Eq.~\eqref{H3}
for a system of size $L=36$ (12 unit cells). The different
symbols denote the different sectors in the CFT description of this critical point, and
the dotted lines indicate the energies predicted by CFT. The numbers indicate the
degeneracies of the states. These degeneracies are in one-to-one correspondence
with the $so(3)_1$ CFT predictions. Here $K=\frac{2\pi k}{L/3}$.}
\label{fig:spectrumn3}
\end{figure}

In Fig.~\ref{fig:spectrumn3} we display the low-lying part of the spectrum for system size
$L = 36$, which corresponds to 12 unit cells. The energies are shifted such
that the ground state has zero energy, and subsequently rescaled such the first excited
state has energy $2 h_{\sigma} = 3/8$, as predicted by the CFT description. With this shift and rescaling, all
the other energies are fixed.  Comparing the energy levels and their degeneracies against the CFT prediction explained in detail
in Appendices~\ref{sec:cfttospectrum} and \ref{sec:characters}, we find excellent agreement with degeneracies matching exactly. This 
convincingly shows that the critical chain Eq.~\eqref{H3} is indeed described by the $so(3)_1 \simeq su(2)_2$ CFT.

\subsection{The $N=4$ case}

For the $N=4$ case we employ the duality transformations (here $0 \leq i,j \leq L/4-1$)
\begin{align}
\label{Transform4}
\sigma^{z}_{4j} &= \tau^{x}_{4j} \tau^{z}_{4j+1} \tau^{z}_{4j+2} \tau^{x}_{4j+3}
\\
\sigma^{z}_{4j+1} &= \tau^{y}_{4j+1} \tau^{y}_{4j+2} 
\nonumber \\
\sigma^{z}_{4j+2} &= \tau^{x}_{4j+1} \tau^{x}_{4j+2}
\nonumber \\
\sigma^{z}_{4j+3} &= \tau^{y}_{4j} \tau^{z}_{4j+1} \tau^{z}_{4j+2} \tau^{y}_{4j+3}
\nonumber \\
\sigma^{x}_{4j} &=
\tau^{y}_{4j} \bigl( \prod_{i<j} \tau^{y}_{4i}  \tau^{y}_{4i+3} \bigr) \mathcal{P}_1
\nonumber \\
\sigma^{x}_{4j+1} &=
\tau^{z}_{4j} \tau^{x}_{4j+1}
\bigl( \prod_{i<j} \tau^{z}_{4i} \tau^{x}_{4i+1} \tau^{x}_{4i+2} \tau^{z}_{4i+3} \bigr)
\nonumber \\
\sigma^{x}_{4j+2} &=
\tau^{y}_{4j+2} \tau^{z}_{4j+3}
\bigl (\prod_{i>j} \tau^{z}_{4i} \tau^{y}_{4i+1} \tau^{y}_{4i+2} \tau^{z}_{4i+3} \bigr)
\nonumber \\
\sigma^{x}_{4j+3} &=
\tau^{x}_{4j+3} \bigl( \prod_{i>j} \tau^{x}_{4i}  \tau^{x}_{4i+3} \bigr) \mathcal{P}_2.
\nonumber
\end{align}
The $\mathcal{P}_n=\prod_{j=0}^{L/4-1}\sigma^z_{4j+n}$ operators appearing in the expressions for
$\sigma^x_{4j}$ and $\sigma^x_{4j+3}$ are included ensure correct commutation relations. However,  they do not affect the form of the 
transformed Hamiltonian.

Applying these transformations to Eq.~\eqref{HN}, we obtain the local Hamiltonian
\begin{equation}
\label{H4}
\begin{split}
H_{so(4)_1} = \sum_{j=0}^{L/2-1} &
\tau^{x}_{2j} \tau^{z}_{2j+1} \tau^{z}_{2j+2} \tau^{x}_{2j+3} + \tau^{x}_{2j+1} \tau^{x}_{2j+2}  + \\ 
& \tau^{y}_{2j} \tau^{z}_{2j+1} \tau^{z}_{2j+2} \tau^{y}_{2j+3} + \tau^{y}_{2j+1} \tau^{y}_{2j+2} \ ,
\end{split}
\end{equation}
which is again translationally invariant. However, unlike in the $N=3$ case where the unit cell contained $N$ sites, here the unit cell has 
the size of $N/2$ sites. This Hamiltonian can also be obtained by applying the condensation framework to two decoupled XY models, as we 
showed in Ref.~\onlinecite{Mansson13}.

\begin{figure}[t]
\includegraphics[width=\columnwidth]{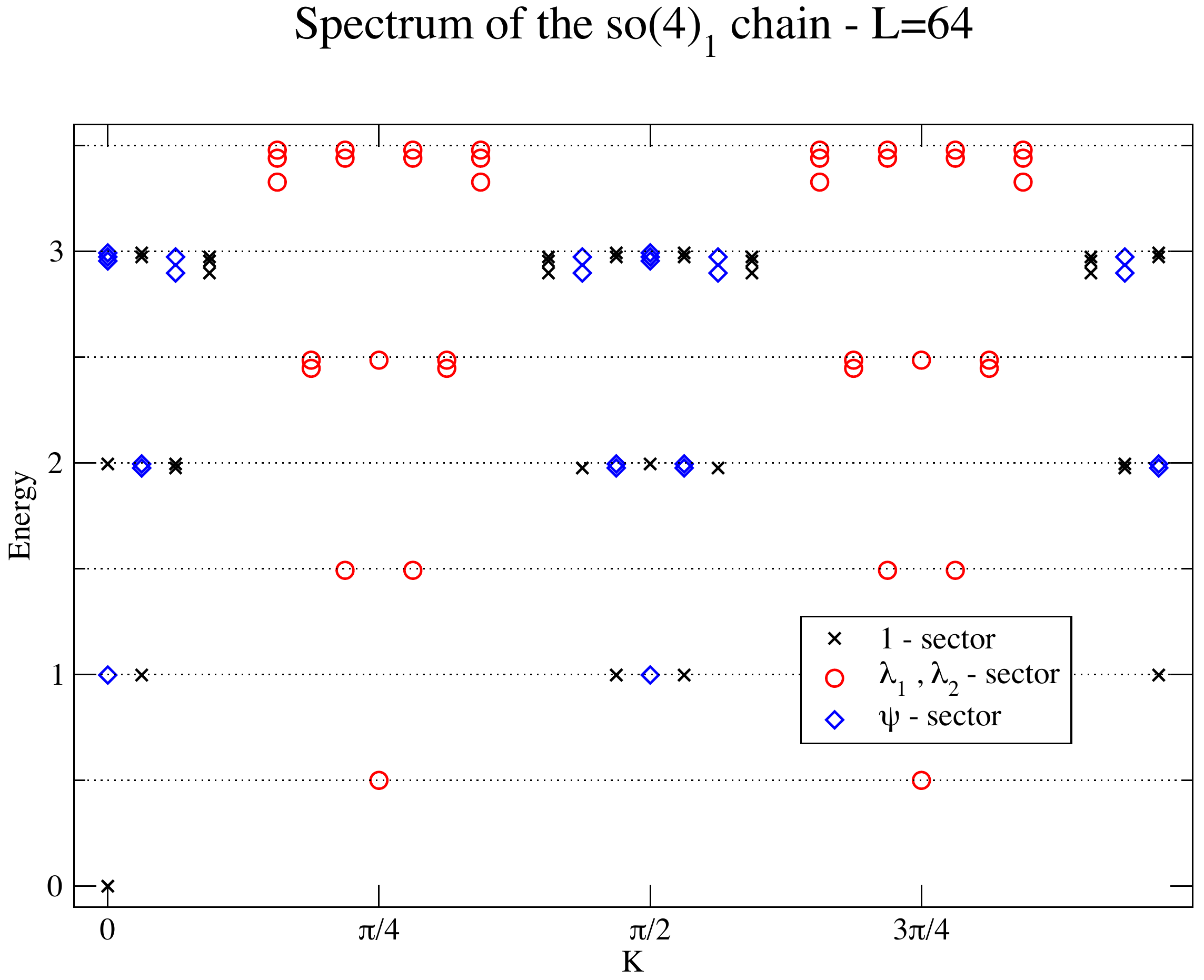}
\caption{The low-lying part of the spectrum of the critical spin chain $H_{so(4)_1}$,
Eq.~\eqref{H4} for system of size $L=64$, or 32 unit cells. The different symbols denote the different sectors in the CFT description of this critical point, and the dotted lines indicate the energies predicted by CFT. While we did not indicate the degeneracies of the states to avoid cluttering the figure, we have verified that they are in exact agreement with the CFT predictions. Here $K=\frac{2\pi k}{L/4}$.}
\label{fig:spectrumn4}
\end{figure}

The structure of the Hamiltonian Eq.~\eqref{H4} is such that it can be solved straightforwardly
by means of a Jordan-Wigner transformation, in the same way as the XY model was solved
in Ref. \onlinecite{lsm61}. In terms of fermions the Hamiltonian describes two decoupled
fermion chains subject to alternating  nearest and third nearest neighbor tunneling.
Explicitly, we obtain the following form
\begin{align}
\label{H4_JW}
H_{so(4)_1} &= \Bigl(
\sum_{j=0}^{L/2-2} 
2(c^\dagger_{2j} c_{2j+3} +  c^\dagger_{2j+1} c_{2j+2}) + h.c. \Bigr) \nonumber \\ &
-2 \mathcal{P}_{so(4)_1} (c^\dagger_{L-2} c_{1} +  c^\dagger_{L-1} c_{0}) + h.c.
\end{align}
When diagonalized, the spectrum is given in terms of two fermions $c_{0,k}$ and $c_{1,k}$ as
\begin{equation} \label{E4}
H_{so(4)_1} = \sum_{k}
\epsilon_{k} \bigl( 2c^{\dagger}_{0,k} c^{\vphantom{\dagger}}_{0,k} -1 \bigr) + \epsilon_{k} \bigl( 2c^{\dagger}_{1,k} c^{\vphantom{\dagger}}_{1,k} -1 \bigr) \ ,
\end{equation}
with
\begin{equation*}
\epsilon_{k} =
\sqrt{2 + 2 \cos \Bigl( \frac{8 \pi k} {L} \Bigr)}  \ .
\end{equation*}
Once again the momenta $k$ runs over integers ($k=0,1,\ldots, L/2-1$) for odd fermion parity ($\mathcal{P}_{so(4)_1}=-1$) and over 
half-integers ($k=1/2, 3/2, \ldots, L/2-1/2$) for even fermion parity ($\mathcal{P}_{so(4)_1}=1$). In Fig.~\ref{fig:spectrumn4} we display 
the rescaled low-lying part of the spectrum of $H_{so(4)_1}$ for a system of size $L=64$. The energy level spacings in the 
rescaled units as well as the degeneracies are again in exact agreements with the $so(4)_1$ CFT predictions, as explained 
in Appendices~\ref{sec:cfttospectrum} and \ref{sec:characters}.

\subsection{The $N=5$ case}

Because the spin chains for $N=1$ and $N=3$ take a rather different form, it is not immediately
obvious what forms the $so(N)_1$ chains take for arbitrary odd $N$. To illustrate the hierarchical structure, we also give the
Hamiltonian $H_{so(5)_1}$ explicitly. Employing the general transformations given in Appendix~\ref{app:general-transform}, it takes the form
\begin{align}
\label{H5}
H_{so(5)_1} &= \\
& \sum_{j=0}^{L/5-1}
 \tau^y_{5j} \tau^z_{5j+1} \tau^z_{5j+2} \tau^z_{5j+3} \tau^y_{3j+4} 
 + \tau^x_{3j+4} \tau^x_{3j+5}  \nonumber \\
& + \tau^x_{5j+1} \tau^z_{5j+2} \tau^z_{5j+3} \tau^z_{5j+4} \tau^y_{3j+5} 
 + \tau^x_{5j+5} \tau^y_{5j+6} \nonumber \\
& + \tau^y_{5j+2} \tau^z_{5j+3} \tau^z_{5j+4} \tau^z_{5j+5} \tau^x_{3j+6} 
 + \tau^y_{5j+6} \tau^x_{5j+7} \nonumber \\
 & + \tau^x_{5j+3} \tau^z_{5j+4} \tau^z_{5j+5} \tau^z_{5j+6} \tau^y_{3j+7} 
 + \tau^x_{5j+7} \tau^y_{5j+8} \nonumber \\
& + \tau^y_{5j+4} \tau^z_{5j+5} \tau^z_{5j+6} \tau^z_{5j+7} \tau^x_{3j+8} 
 + \tau^y_{5j+8} \tau^x_{5j+9} \nonumber \ .
 \end{align}
Like the $N=3$ case, also this Hamiltonian can be brought to a form similar to \rf{H3_2} where the translational symmetry is manifestly broken only by the sign of one of the 5-spin terms. To bring this model in a diagonal form, one uses exactly the same steps as
for $N=3$, with the following result
\begin{align}
\label{H5E}
H_{so(5)_1} & = \sum_{k}\sum_{n=0}^{4}
\epsilon_{n,k}
\bigl( 2c^{\dagger}_{n,k} c^{\vphantom{\dagger}}_{n,k} -1 \bigr) \\
\nonumber
\epsilon_{n,k} &=
\begin{cases}
\sqrt{2 + 2 \cos \bigl( \frac{2\pi k N}{L}\bigr)} & \qquad \text{for } n = 0 \\
\sqrt{2 - 2 \cos \bigl( \frac{2\pi k N}{L}\bigr)} & \qquad \text{for } n=1,2,3,4 \ .
\end{cases}
\end{align}

\subsection{The general $N$ case}

The examples above are part of a hierarchy of exactly solvable spin-1/2 models with $so(N)_1$ critical behavior. Their microscopic 
structure depends systematically on $N$, such that even and odd $N$ cases form different sets of models. We present in 
Appendix~\ref{app:general-transform} the most general duality transformations to bring the Hamiltonians $H_{so(N)_1}$ into a 
local and translationally invariant form. Here we present these Hamiltonians and their solutions for all $N$.

\subsubsection{Odd $N$}
The structure of the Hamiltonians for odd $N$ is as follows. There are pairs of 2-spin and
$N$-spin operators. One pair will always consists of an $N$-spin term acting on $N$ adjacent spins,
with two $\tau^y$ operators straddling a string of $N-2$ $\tau^z$ operators. The Jordan-Wigner solvability of terms of this form was first pointed out by Suzuki \cite{suzuki71b}. The second term of this pair is a product of two adjacent $\tau^x$ operators. These two terms do not
commute with one another (but they commute with all other terms in the Hamiltonian).
The remaining $N-1$ pairs consists of an $N$-spin term, where an $\tau^x$ and a
$\tau^y$ straddle a string of $N-2$ $\tau^z$'s, while the second term is the product of neighboring
$\tau^x$ and a $\tau^y$. Again, the two members of a pair do not commute, while they
commute with all other terms in the Hamiltonian. Because of the structure with the strings
of $\tau^z$ operators, these models can be solved by means of a Jordan-Wigner
transformation. Explicitly, the Hamiltonian for general odd $N$ reads
\begin{widetext} 
\begin{equation} \label{HNodd}
\begin{split}
&H_{so(N)_1} = \sum_{j=0}^{L/N-1} \Bigl(
\tau^{y}_{jN} \tau^{z}_{jN+1} \cdots \tau^{z}_{jN+N-2} \tau^{y}_{jN+N-1} +
\tau^{x}_{jN+N-1} \tau^{x}_{jN+N} + \\ &
\sum_{n=1}^{(N-1)/2}
\tau^{x}_{jN+(2n-1)} \tau^{z}_{jN+(2n-1)+1} \cdots
\tau^{z}_{jN+(2n-1)+(N-2)} \tau^{y}_{jN+(2n-1)+(N-1)} +
\tau^{x}_{jN+(2n-1)+(N-1)} \tau^{y}_{jN+(2n-1)+N} + \\ &
\sum_{n=1}^{(N-1)/2}
\tau^{y}_{jN+2n} \tau^{z}_{jN+2n+1} \cdots
\tau^{z}_{jN+2n+(N-2)} \tau^{x}_{jN+2n+(N-1)} +
\tau^{y}_{jN+2n+(N-1)} \tau^{x}_{jN+2n+N} \Bigr)\ .
\end{split}
\end{equation}
\end{widetext}
As was the case for $N=3$, this most general spin chain can also be diagonalized with a Jordan Wigner transformation. The unit cell has 
the size of $N$ sites, which implies that  the spectrum is given in terms of $N$ fermionic operators $c_{n,k}$. For all odd $N$ it is given by
\begin{align}
H_{so(N)_1} &= \sum_{k}\sum_{n=0}^{N-1}
\epsilon_{n,k}
\bigl( 2c^{\dagger}_{n,k} c^{\vphantom{\dagger}}_{n,k} -1 \bigr) \\ \nonumber
\epsilon_{n,k} &=
\begin{cases}
\sqrt{2 + 2 \cos \bigl( \frac{2\pi k N}{L}\bigr)} & \qquad \text{for } n = 0 \\
\sqrt{2 - 2 \cos \bigl( \frac{2\pi k N}{L}\bigr)} & \qquad \text{for } 1\leq n \leq N-1 \ .
\end{cases}
\end{align}
As we discussed with the explicit examples above, the momenta $k$ run in the usual way over integers or half-integers depending 
on the total fermion parity \rf{PN}.

\subsubsection{Even $N$}

The general even $N$ Hamiltonians consist of $N/2$ pairs of terms,
where one member of a pair consists of two $\tau^x$ operators straddling a  
string of $\tau^z$ operators, while the other member of the pair consists of two
$\tau^y$ operators straddling a string of $\tau^z$ operators.
The general Hamiltonian for even $N$ reads
\begin{align}
\label{HNeven}
& H_{so(N)_1} = \sum_{j=0}^{2(L/N)-1}\sum_{n = 0}^{N/2-1} \\ \nonumber
& \tau^{x}_{j N/2 + n}
\tau^{z}_{j N/2 + n + 1} \cdots
\tau^{z}_{j N/2 + N - n -2}
\tau^{x}_{j N/2 + N - n -1} +
\\ \nonumber &
\tau^{y}_{j N/2 + n}
\tau^{z}_{j N/2 + n + 1} \cdots
\tau^{z}_{j N/2 + N - n -2}
\tau^{y}_{j N/2 + N - n -1} 
\end{align}
As the unit cell for these models contains always only $N/2$ sites, the spectrum is given in terms of $N/2$ fermions $c_{n,k}$. Explicitly,
\begin{align}
H_{so(N)_1} &= \sum_{k}\sum_{n=0}^{N/2-1}
\epsilon_{k}
\bigl( 2c^{\dagger}_{n,k} c^{\vphantom{\dagger}}_{n,k} -1 \bigr) \\ \nonumber
\epsilon_{k} &= \sqrt{2 + 2 \cos \Bigl( \frac{2 \pi k N} {L} \Bigr)} \ ,
\end{align}
where the momenta $k$ again depends in the usual way on the fermionic parity \rf{PN}.

We have performed checks on the low-lying states for all these
Hamiltonians up to $N=16$ against the $so(N)_1$  CFT predictions presented in  Appendices~\ref{sec:cfttospectrum} and \ref{sec:characters}. Labeling of the states by the primary fields and their low-lying descendants as well as the expected degeneracies are found to be in complete agreement with the CFT predictions.

\section{The structure of the phase diagrams of the $so(N)_1$ models.}
\label{sec:phase-diagram}

Our construction was motivated by the condensation framework that related different CFTs that describe the criticality of different spin chains. Therefore, the hierarchy of spin chains we constructed are all fine-tuned to a critical point. However, as is well known, the two simplest members of this hierarchy, 
the TFI chain ($N=1$) and the XY chain ($N=2$), are typically gapped with the critical points separating different gapped phases. 
The situation is very similar for the models for general $N$. In these models the number of different types of terms grows 
linearly with $N$, so in principle one can introduce many different coupling constants for which the models can still be 
solved exactly. 

\subsection{Phase diagrams of odd $N$ chains}

We start by considering the odd $N$ chains that contain two-spin and $N$-spin terms. For concreteness we consider the case $N=3$ and introduce the most general
coupling parameters
\begin{align}
\label{H3gen}
\hat{H}_{so(3)_1}= \sum_j \Bigl( & g_{0} \tau^x_{3j+2} \tau^x_{3j+3} + h_{0}  \tau^y_{3j} \tau^z_{3j+1} \tau^y_{3j+2} 
\\ \nonumber &
+ g_{1} \tau^x_{3j+3} \tau^y_{3j+4} + h_{1} \tau^x_{3j+1} \tau^z_{3j+2} \tau^y_{3j+3}
\\ \nonumber &
+ g_{2} \tau^y_{3j+4} \tau^x_{3j+5} + h_{2} \tau^y_{3j+2} \tau^z_{3j+3} \tau^x_{3j+4} \Bigr).
\end{align}
Diagonalizing this model gives again a spectrum of three fermions \rf{H3E}, but now with the generic dispersion relations
\begin{equation*}
\epsilon_{n,k} =
\begin{cases}
\sqrt{g_{0}^2 + h_{0}^2 + 2 g_{0} h_{0} \cos \bigl( \frac{2\pi k}{(L/3)}\bigr)}
& \;\; (n = 0) \\
\sqrt{g_{n}^2 + h_{n}^2 - 2 g_{n} h_{n} \cos \bigl( \frac{2\pi k}{(L/3)}\bigr)}
& \;\; (n = 1, 2) \ .
\end{cases}
\end{equation*}

\begin{figure}[t]
\includegraphics[width=0.8\columnwidth]{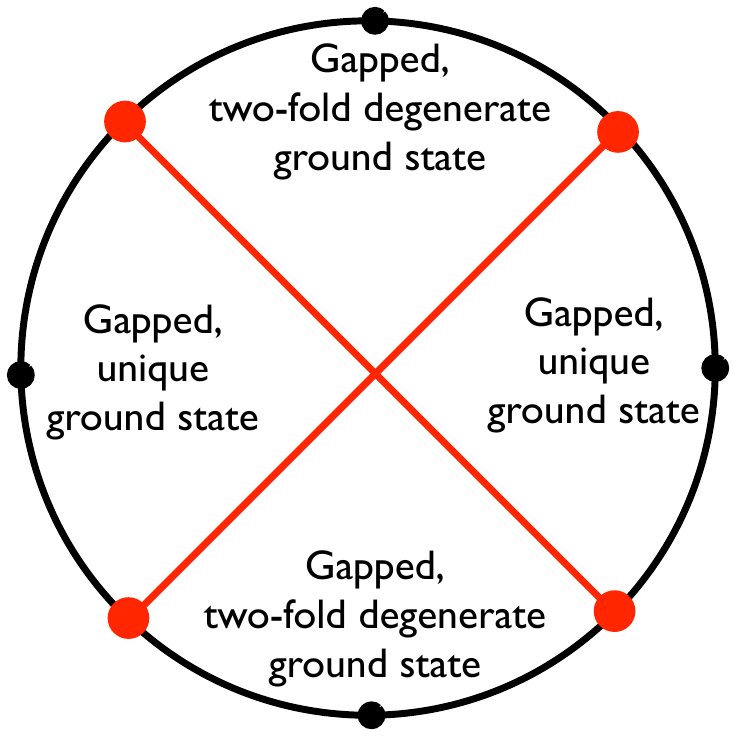}
\caption{The schematic phase diagram of $\hat{H}_{so(3)_1}$, Eq.~\eqref{H3gen}, as a function of the
angle $\theta$, defined as $\cos\theta = g = g_{0} = g_{1} = g_{2}$
and $\sin\theta = h = h_{0} = h_{1} = h_{2}$. The critical points at $|g|=|h|$ are indicated
by the red dots, while the black dots represent the special points with either $g=0$ or $h=0$ where all terms in the
Hamiltonian commute with one another.}
\label{fig:phase-diagram}
\end{figure}

Let us first restrict to analyze the phase diagram when the relative couplings between the two spin terms
(i.e., the nearest neighbor couplings in the fermionic version) and the $N$-spin terms ($(N-1)^{\rm th}$ nearest neighbor coupling for fermions) are varied. To this end we set $g_{0} = g_{1} = g_{2} = g$ and $h_{0} = h_{1} = h_{2} = h$. Fig.~\ref{fig:phase-diagram} shows that similar to the TFI chain, the $N=3$ chain is critical only when $|g| = |h|$, with a gap opening up immediately when one moves 
away from these four points. The ground states in the four gapped phases can be characterized by considering the special 
points $g = 0, \ h = \pm 1$ and $g = \pm 1, \ h = 0$, where all terms in the Hamiltonian commute with each other. When $g = 0$ we find 
a unique, `polarized' ground state. The `polarization' of the spin depends on the position of the spins in the unit cell, and can be obtained
from the Hamiltonian in a straightforward way. For $h = 0$ the system has a two-fold degenerate ground state in which neighboring spin are 
`aligned'. Again, in which direction the spins are aligned depends on the position of the spins in the unit cell.

These phases are the direct analogs of the gapped phases present in the TFI model, with the three-spin terms playing 
the role of the magnetic field term. In its fermionic representation the TFI chain realizes Kitaev's $p$-wave paired nanowire,\cite{Kitaev01} where the phase with degenerate ground state corresponds to the weak pairing topological phase and the 'polarized' phase to the topologically trivial strong pairing phase. For open boundary conditions the first hosts localized Majorana modes at the chain ends. As the odd $N$ models of our hierarchy exhibit similar phase diagrams, and the classification of topological phases admits only two topologically distinct phases for particle-hole symmetric models in one spatial dimension\cite{Kitaev09,Schnyder09}, we also expect similar behavior from the gapped phases of our hierarchy models.  

For general odd $N$ and general coupling constants $g_n$, and $h_n$, with $n=0,1,\ldots N-1$, the situation is as follows.
As long as $|g_n| \neq |h_n|$ for {\em all} $n$, the system has a gap, but as soon as for an arbitrary value of $n$ one has $|g_n| = |h_n|$,
the gap closes. This implies that the phase diagram is in general richer than that shown in Fig.~\ref{fig:phase-diagram}, that is valid only for $g=g_n$ and $h=h_n$ for all $n$. We leave the detailed study of the full phase diagrams for future work.

\subsection{Phase diagrams of even $N$ chains}

Like all the odd $N$ models are generalizations of the TFI chain, so can all the even $N$ models be viewed as generalizations of the XY chain. 
The general Hamiltonian \rf{HNeven} implies that even $N$ models always contain 2-, 4-,$\ldots$ and $N$-spin terms.
Focusing on the simplest case $N=4$, we introduce again general couplings for which the Hamiltonian reads
\begin{align}
\label{H4gen}
\nonumber
\hat{H}_{so(4)_1} = \sum_{j=0}^{L/2-1} &
 g_{0} \tau^{x}_{2j} \tau^{z}_{2j+1} \tau^{z}_{2j+2} \tau^{x}_{2j+3} + h_{0} \tau^{y}_{2j+3} \tau^{y}_{2j+4} + \\ 
& g_{1} \tau^{x}_{2j+1} \tau^{x}_{2j+2} + h_{1} \tau^{y}_{2j+2} \tau^{z}_{2j+3} \tau^{z}_{2j+4} \tau^{y}_{2j+5} \ .
\end{align}
We have labeled the terms such that non-commuting terms have coupling constants with the
same index. For general couplings the fermionized Hamiltonian also has pairing terms, which means that after diagonalization, the spectrum is given in terms of two fermions as in
Eq.~\eqref{E4}, but with more complicated dispersions for the fermions. For $n=0,1$, we find
\begin{equation*}
\epsilon_{n,k} =
\sqrt{g_{n}^2 + h_{n}^2 + 2 g_{n} h_{n} \cos \bigl( \frac{8 \pi k}{L}\bigr)} \ .
\end{equation*}

The phase diagram for the case $g_0 = g_1 = g$ and $h_0 = h_1 = h$ has qualitatively the same structure as the phase diagram of the
XY model. The difference to the phase diagrams of the odd $N$ chains (see Fig.~\ref{fig:phase-diagram}), is that now the four critical points at $|g| = |h|$ separate four gapped phases, that all have a unique ground state. Also similar to the odd $N$ cases, all even $N$ chains with arbitrary coupling constants are
gapped as long as for {\em all} values of $n$, one has $|g_n| \neq |h_n|$, and critical otherwise. This means that also the even $N$ cases have phase diagrams that go beyond that of the XY chain. Their systematic study certainly warrants further investigation.

\section{Beyond exact solvability: The spin-1 Blume-Capel model}
\label{sec:blume-capel}

In the previous sections, we used the condensation picture to construct $so(N)_1$ critical
spin chains starting from $N$ decoupled TFI chains. The constructed models are solvable via a Jordan-Wigner transformation. In this section, we argue that this condensation picture based construction is general and applies also to critical chains that are not exactly solvable. 

\subsection{The Blume-Capel model}

To show this, we consider the so-called Blume-Capel model, which is a spin-1 model, exhibiting a tri-critical point described by the
tri-critical Ising CFT. In its two-dimensional classical incarnation, the Blume-Capel
model is an Ising model with vacancies \cite{blume66,capel66}. In its one-dimensional
quantum version we consider here, it takes the form of a spin-1 model,
exhibiting an interesting phase diagram, see for instance Ref.~\onlinecite{vongehlen90}.
The Hamiltonian of the one-dimensional $L$ site quantum Blume-Capel model is given by\cite{footnote2}
\begin{equation} \label{HBC}
H_{\rm BC} = \sum_{j=0}^{L-1} -S^{x}_j S^{x}_{j+1} + \alpha \bigl( S^{x}_j \bigr)^2
+ \beta  S^{z}_j  \ .
\end{equation}
We use the standard representation for the spin-1 matrices, namely
\begin{align}
S^{x}_{i} &= \frac{1}{\sqrt{2}}
\begin{pmatrix} 0 & 1 & 0 \\ 1 & 0 & 1 \\ 0 & 1 & 0 \end{pmatrix} &
S^{z}_{i} &=
\begin{pmatrix} 1 & 0 & 0 \\ 0 & 0 & 0 \\ 0 & 0 & -1 \end{pmatrix} 
\nonumber \\
S^{y}_{i} &= \frac{i}{\sqrt{2}}
\begin{pmatrix} 0 & -1 & 0 \\ 1 & 0 & -1 \\ 0 & 1 & 0 \end{pmatrix}
\ .
\end{align}
As we did for the $so(N)_1$ models, we also assume periodic boundary conditions throughout.
To get an idea about the phase diagram of the model,
we first consider the case $\alpha = 0$. Then, the model is the
spin-1 version of the transverse field Ising model, and exhibits a second order phase
transition (at $\beta = \sqrt{2}$) in the Ising universality class,
just as the spin-1/2 transverse field Ising model.
In the case that $\beta = 0$, the model exhibits a first-order phase transition (at $\alpha =1$).
For arbitrary $\alpha$ and $\beta$, the phase transitions mentioned above are actually
lines of phase transitions, which meet at a tri-critical point located at
$\alpha \approx 0.910207$ and $\beta \approx 0.415685$. We refer to Ref.~\onlinecite{vongehlen90} for more details on the phase diagram.

At the tri-critical point, the system is described by the tri-critical Ising CFT. This CFT has central charge $c=7/10$ and it is the second model in a series of CFTs, called `minimal models',
where the Ising CFT is the first in the series \cite{bpz84}. It contains six primary fields, which we label as
$\{ \id, \sigma, \sigma',\epsilon,\epsilon',\epsilon''\}$, with scaling dimensions
$h_{\id} = 0$, $h_{\epsilon''} = 3/2$,
$h_{\epsilon} = 1/10$, $h_{\epsilon'} = 3/5$,
$h_{\sigma} = 3/80$ and $h_{\sigma'} = 7/16$.
To describe the fusion rules
of the tri-critical Ising CFT, it is easiest to give each fields two labels,
one label representing a set of particles $\{\id,\sigma,\psi\}$
satisfying the Ising fusion rules, the other representing the so-called
Fibonacci anyon model, $\{\id,\tau\}$, with the only non-trivial fusion rule
$\tau\times\tau = \id + \tau$. Using the correspondence
$\id = (\id,\id)$, $\epsilon'' = (\psi,\id)$,
$\epsilon = (\psi,\tau)$, $\epsilon' = (\id,\tau)$, 
$\sigma = (\sigma,\tau)$ and $\sigma' = (\sigma,\id)$, one can derive the fusion
rules of the tri-critical Ising CFT. For instance, we have
$\sigma \times \sigma = (\sigma,\tau)\times (\sigma,\tau) =
(\id,\id) + (\id,\tau) + (\psi,\id) + (\psi,\tau) = \id + \epsilon' + \epsilon'' + \epsilon$.

\begin{figure}[t]
\includegraphics[width=\columnwidth]{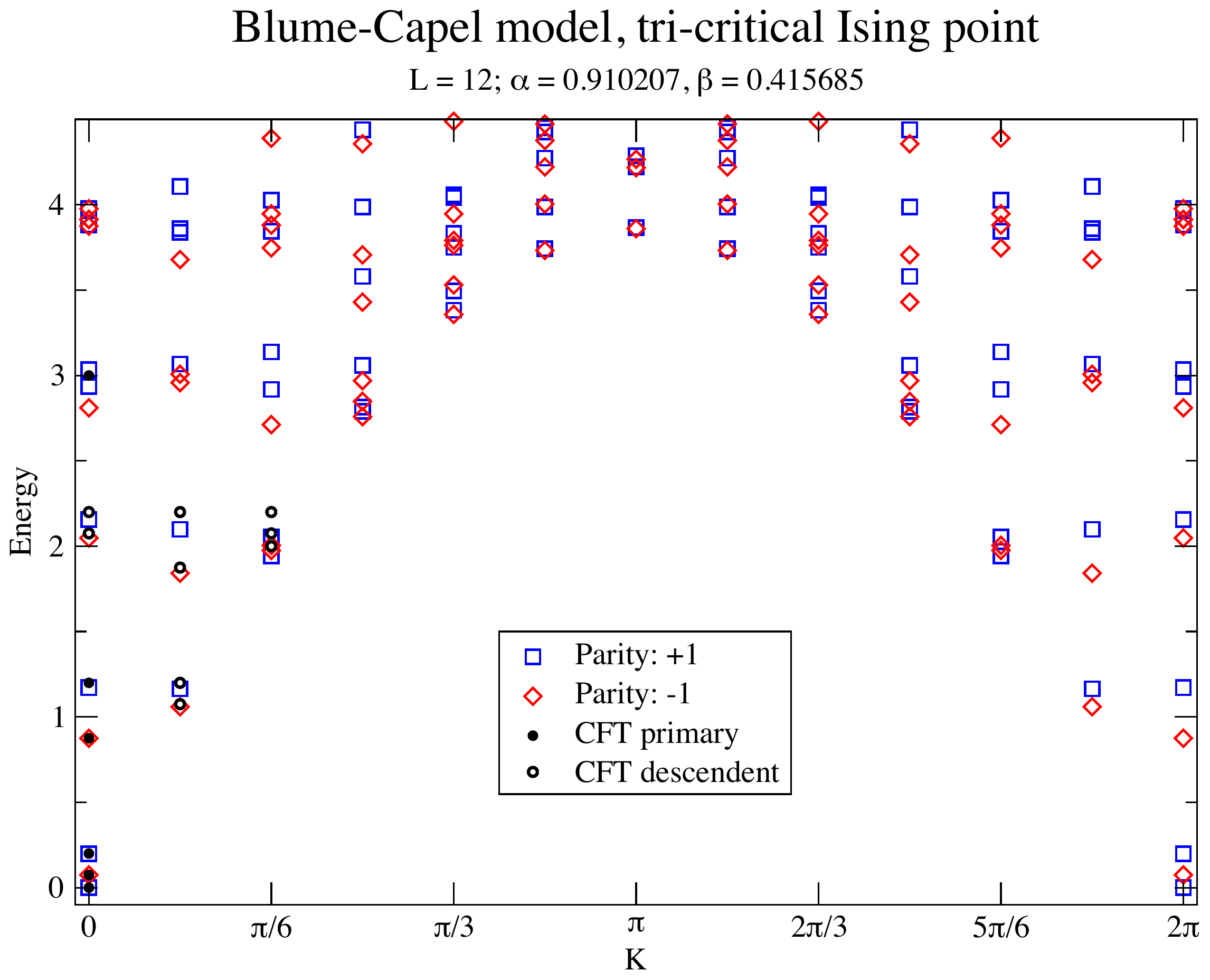}
\caption{The low-lying part of the spectrum of the Blume-Capel chain at the tri-critical point.
The blue squares (red diamonds) indicate the states with $\mathcal{P}^z$ eigenvalues $+1$
($-1$). The black dots (blue squares) indicate the position of the primary (descendent) fields
of the tri-critical Ising CFT.}
\label{fig:bcspectrum}
\end{figure}

The field $\epsilon''$ has scaling dimension $h_{\epsilon''} = 3/2$, which means
that it is a fermionic field. Thus, if we consider a
Blume-Capel model with (apart from the on-site terms) next-nearest neighbor interactions
only, we obtain two decoupled chains, whose criticality is described by doubled tri-critical Ising CFT that contains a bosonic field with $h_{\epsilon'',\epsilon''}=3$. This suggests that it could be possible to
add an appropriate condensing boundary term, such that we obtain a different spin-1 chain, whose criticality is related to that of the doubled Blume-Capel model via the condensation framework. To do this one needs to find a symmetry operator that separates the confined and non-confined CFT sectors. Following Kennedy and Tasaki \cite{kennedy92}, we introduce the following operators
\begin{align}
P^{x} &= -e^{i \pi S^x} = 2\bigl(S^x\bigr)^2 - \id =
\begin{pmatrix}
0 & 0 & 1 \\
0 & 1 & 0 \\
1 & 0 & 0 \\
\end{pmatrix} \\ \nonumber
P^{y} &= -e^{i \pi S^y} = 2\bigl(S^y\bigr)^2 - \id =
\begin{pmatrix}
0 & 0 & -1 \\
0 & 1 & 0 \\
-1 & 0 & 0 \\
\end{pmatrix} \\ \nonumber
P^{z} &= -e^{i \pi S^x} = 2\bigl(S^x\bigr)^2 - \id =
\begin{pmatrix}
1 & 0 & 0 \\
0 & -1 & 0 \\
0 & 0 & 1 \\
\end{pmatrix} \ .
\end{align}
The operator $\mathcal{P}^{z} = \prod_{i=0}^{L-1} P^{z}_i$
commutes with \rf{HBC}
for all values of $\alpha$ and $\beta$.
States with an even (odd) number of sites with $S^z_i=0$ have $\mathcal{P}^{z}$
eigenvalue $+1$ ($-1$). In Fig.~\ref{fig:bcspectrum}, we give the low-lying spectrum of the Blume-Capel Hamiltonian,
indicating the $\mathcal{P}^{z}$ eigenvalues of all the states. We find that the states deriving from the CFT sectors
$\id, \epsilon, \epsilon', \epsilon''$ have $\mathcal{P}^z$ eigenvalue $+1$, while
the states deriving from $\sigma,\sigma'$ have $\mathcal{P}^z$ eigenvalue $-1$.
This division of the states into two groups is similar to the TFI chain, suggesting that a condensing boundary term could be constructed from this operator. The spectrum also confirms that the tri-critical point is indeed described by the tri-critical Ising CFT.

\subsection{Constructing a new spin-1 chain from condensation framework}

We now use the correspondence between the symmetry sectors $\mathcal{P}^{z}=\pm1$ and the tri-critical Ising CFT primary fields to construct a condensing boundary term for the Blume-Capel chain with next-nearest neighbor interactions only.

The CFT describing the tri-critical point of the decoupled system is given by
the product of two tri-critical Ising CFTs. This CFT has central charge $c=7/5$ and it contains 36 primary fields. The field labeled by $(\epsilon'',\epsilon'')$ has scaling dimension $h_{(\epsilon'',\epsilon'')}=3$, which means that it is a boson that can condense. Without going into details, the
construction of the theory after condensing this boson follows the same lines as the
case of the ${\rm Ising}^{\times 2}$ theory, by making use of the fact that the fusion rules of the tri-critical Ising
CFT can be seen as a `product' of an Ising and a Fibonacci sector.
In the end one obtains a theory containing the 16 fields
\begin{align}
\nonumber
\id &= (\id,\id) & \epsilon'' &= (\id,\epsilon'') &
\sigma_1 &= (\sigma,\sigma)_1 & \sigma'_1 &= (\sigma',\sigma)_1
\\ \nonumber
\epsilon &= (\epsilon,\epsilon) & \epsilon' &= (\epsilon,\epsilon') &
\sigma_2 &= (\sigma,\sigma)_2 & \sigma'_2 &= (\sigma',\sigma)_2 \\ \nonumber
\epsilon_l &= (\epsilon,\id) & \epsilon'_l &= (\epsilon',\id) &
\sigma''_1 &= (\sigma',\sigma')_1 & \sigma'_3 &= (\sigma,\sigma')_2 \\
\epsilon_r &= (\id,\epsilon) &\epsilon'_r &= (\id,\epsilon') &
\sigma''_2 &= (\sigma',\sigma')_2 & \sigma'_4 &= (\sigma,\sigma')_2
\label{eq:bc-cond-fields}
\end{align}
The fusion rules for these fields are equivalent to the
fusion rules of ${\rm Fib} \times {\rm Fib} \times Z_4$, where ${\rm Fib}$ stands for the
Fibonacci fusion rules introduced above, and $Z_4$ correspond to charges $l=0,1,2,3$, which
upon fusion are added modulo 4 (i.e., they correspond to the fusion rules of $u(1)_4$).

The condensation process confines all the primary fields that contain only a single $\sigma$ or $\sigma'$ field. In the Blume-Capel chain with next nearest neighbor interactions all states labeled by these fields reside in the symmetry sectors for which $\mathcal{P}^z_{\rm even}=-\mathcal{P}^z_{\rm odd}$, where $\mathcal{P}^z_{\rm even}= \prod_{j, {\rm even}} P^{z}_j$ and $\mathcal{P}^z_{\rm odd}= \prod_{j, {\rm odd}} P^{z}_j$ are the symmetry operators for the two decoupled chains on even and odd sites, respectively. The lack of an exact solution means that there is now no obvious correspondence between the fermion boundary conditions and the symmetry sectors, but the similar spectral partitioning as in the system of two TFI chains motivates us still to construct a similar condensing boundary term.  In precise analogy to \rf{HB2}, this term is given by
\begin{equation}
\label{eq;bc-with-boundary}
\begin{split}
H_{BC}^B = (\id - \mathcal{P}^z_{\rm even}) S^{x}_{L-1}S^{x}_{1} + (\id - \mathcal{P}^z_{\rm odd}) S^{x}_{L-2}S^{x}_{0} \ .
\end{split}
\end{equation}

\begin{figure}[t]
\includegraphics[width=\columnwidth]{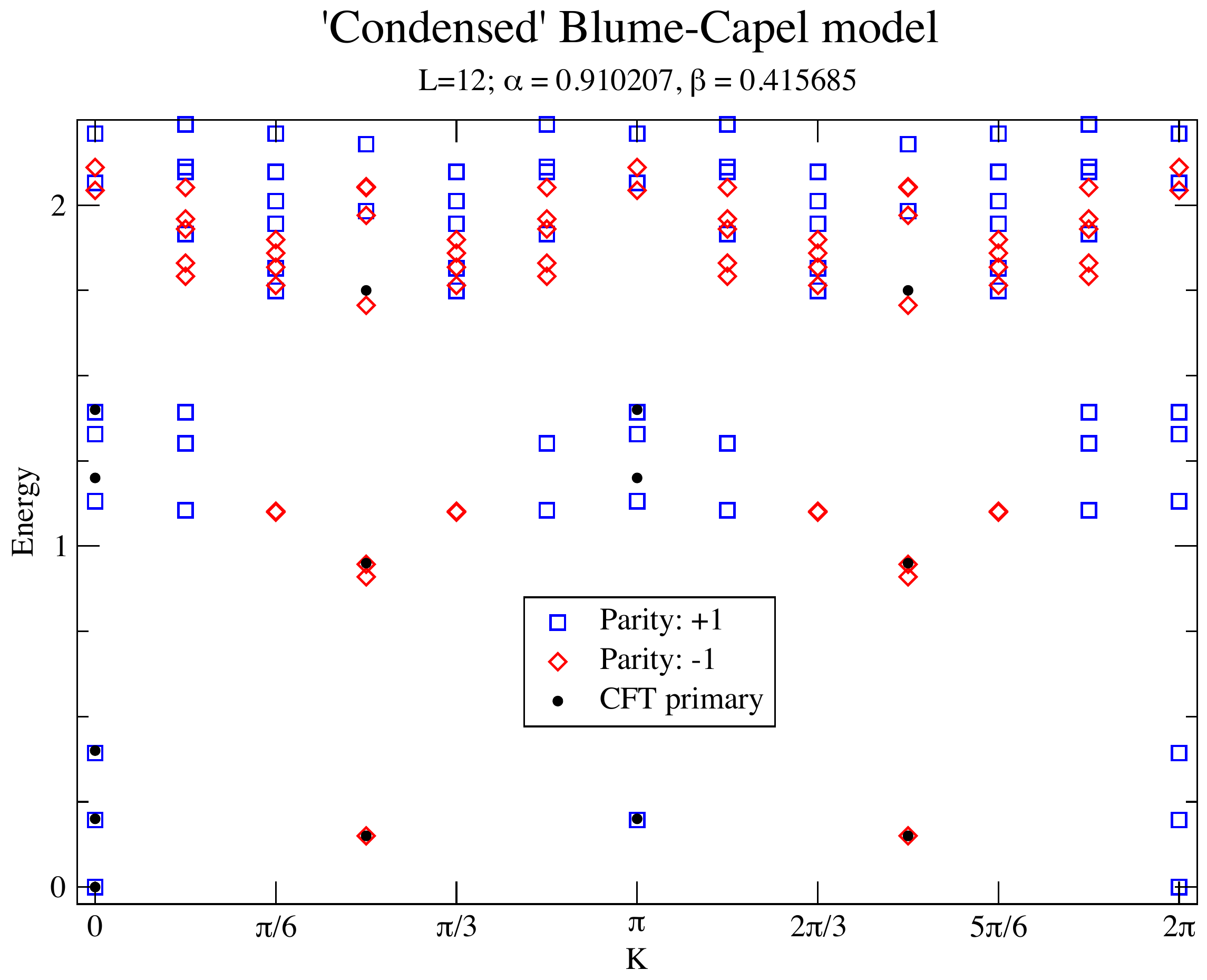}
\caption{The low-lying part of the spectrum of the model Eq.~\eqref{eq:bc-cond-ham}
at the tri-critical point.
The blue squares (red diamonds) indicate the states with $\mathcal{P}^z$ eigenvalues $+1$
($-1$). The black dots indicate the position of the primary fields \rf{eq:bc-cond-fields}  of the CFT describing
this critical point.}
\label{fig:bc-cond-spectrum}
\end{figure}

Just as was the case for the spin-1/2 chains we considered, it is possible to perform a duality
transformation on the spin-1 operators and transform $H_{\rm BC}+H_{\rm BC}^B$ into a translationally invariant form.
This transformations can be compactly written as $H^{\rm cond}_{\rm BC} = U (H_{\rm BC}+H_{\rm BC}^B) U^\dagger$, where
\be
  U = \prod_{\substack{j<k \\ j \, {\rm even}\\ k \, {\rm odd}}} e^{i\pi S^z_j S^z_k}.
\ee
We note that the form of this operator is closely related to the one
considered in Ref.~\onlinecite{oshikawa92}. More explicitly, the individual spin operators transform as
\begin{align}
\nonumber
T^{x}_{j, \rm {even}} &=  \bigl( \prod_{k<j, {\rm odd}} P^{z}_k \bigr) S^{x}_j &
T^{x}_{j, \rm {odd}} &= S^{x}_j \bigl( \prod_{k>j, {\rm even}} P^{z}_k \bigr) \\
\nonumber
T^{y}_{j, \rm {even}} &=  \bigl( \prod_{k<j, {\rm odd}} P^{z}_k \bigr) S^{y}_j &
T^{y}_{j, \rm {odd}} &= S^{y}_j \bigl( \prod_{k>j, {\rm even}} P^{z}_k \bigr) \\
T^{z}_{j, \rm {even}}  &= S^{z}_j &
T^{z}_{j, \rm {odd}}  &= S^{z}_j \ ,
\end{align}
which gives the local and translationally invariant Hamiltonian
\begin{equation}
\label{eq:bc-cond-ham}
H^{\rm cond}_{\rm BC} =
\sum_{j=0}^{L-1} -T^{x}_j T^{x}_{j+2} + 2 T^{x}_j \bigl(T^{z}_{j+1}\bigr)^2 T^{x}_{j+2}
+ \alpha \bigl( T^{x}_j \bigr)^2
+ \beta  T^{z}_j \ .
\end{equation}
We have diagonalized this Hamiltonian exactly for a system of size $L=12$. Fig.~\ref{fig:bc-cond-spectrum} shows that although the finite size effects are rather substantial, the spectrum is in excellent agreement with the predicted CFT with the 16 primary fields are given by \rf{eq:bc-cond-fields}.

To gain insight in the CFT with the primary fields \rf{eq:bc-cond-fields} describing the tri-critical point of the Hamiltonian
Eq.~\eqref{eq:bc-cond-ham}, we note the following. The starting point was the tri-critical
Ising CFT, which is the first in a series of CFTs having $\mathcal{N}=1$
supersymmetry \cite{friedan84}. This series is labeled by the integer $m = 3,4,\ldots$,
with central charges $c = 3/2 (1-8/(m(m+2)))$, giving $c=7/10$ for $m=3$
Condensation transitions conserve the central charge, and after condensing the boson in the product theory of two
tri-critical Ising CFTs, one finds a theory which seems to inherit the $\mathcal{N}=1$ supersymmetry. The central charge of this
theory is $c= 7/5$, which is indeed among the central charges of the supersymmetric
`minimal' models, namely for $m=10$. For $m$ even, the central charge does not
uniquely specify the supersymmetric CFT. There are different CFTs with the same
central charge that form different modular invariants \cite{cappelli87}.
The standard diagonal modular invariant for $m=10$ has 76 (Virasoro) primary fields.
It turns out that the theory at hand corresponds to an exceptional modular invariant,
namely $(D_6,E_6)$, which indeed has 16 primary fields. We verified explicitly that
the CFT we constructed is indeed the $(D_6,E_6)$ invariant \cite{cappelli87}
of the $m=10$ CFT, by making use of the explicit form of the characters of the Virasoro minimal
models \cite{r85} and  the characters of the $\mathcal{N}=1$ supersymmetric
`minimal' models \cite{gko86}.

The qualitative similarity of the condensing boundary term suggests that this counterpart of condensation transition in two Blume-Capel models is just one example of a larger hierarchy, exactly like the coupling two TFI chains by means of such term was the simplest example of the $so(N)_1$ hierarchy. Indeed, since the fusion rules of a tri-critical Ising CFT could be understood as the product Ising $\times$ Fib, and the condensed boson in the doubled theory was formed out of the fermions in the two Ising-like sectors, the transition acts trivially in the two Fib sectors. At the level of fusion rules, the transition modified them as Ising$^{\times 2} \times$ Fib$^{\times 2} \to Z_4 \times$ Fib$^{\times 2}$. As the corresponding CFT contains a fermion in the spectrum \rf{eq:bc-cond-fields}, one can imagine considering a system of three or more tri-critical Blume-Capel models coupled together by a generalized condensing boundary term similar to the one we derived for three TFI chains in Appendix \ref{app:boundary-hamiltonians}. Starting from $N$ Blume-Capel models, one would then expect a transition
\be
  {\rm Ising}^{\times N} \times {\rm Fib}^{\times N} \to so(N)_1 \times {\rm Fib}^{\times N},
\ee
where all the CFTs refer only to the fusion rules (scaling dimensions have to be worked out separately). The resulting criticality would be described by a CFT with central charge $c=7N/10$, but in general these would not correspond to an $\mathcal{N}=1$ supersymmetric `minimal' models, because their central charge is maximally $c=3/2$. We leave it for future work to study whether such hierarchy could be realized at the level of spin-1 chains.

\section{Discussion}
\label{sec:conclusions}

We have generalized the insight of Ref.~\onlinecite{Mansson13}, that two-dimensional condensate-induced transitions\cite{Bais09} have counterparts in critical spin chains, to construct a hierarchy of exactly solvable spin-1/2 chains with $so(N)_1$ critical points. Our construction is based on first coupling together $N$ critical TFI chains by means of a non-local Hamiltonian term -- a condensing boundary term that can be derived from the condensation picture -- and then transforming these coupled systems into a translationally invariant form by means of general spin duality transformations. As our construction respects the symmetries of the decoupled TFI system, the resulting chains are also exactly solvable with a Jordan-Wigner transformation. Comparing the energy spectra of the constructed chains to the predictions by CFT, we explicitly verified that all the chains in our hierarchy are indeed critical and described by $so(N)_1$ CFTs. While the exact solvability is an attractive feature of our hierarchy, the condensation transition motivated approach goes beyond exactly solvable models. We showed that a similar strategy could be applied also to two decoupled tri-critical Blume-Capel models, TFI-like spin-1 models with to our understanding no known exact solution, and derived another critical spin-1 chain with a critical point described by the predicted supersymmetric minimal model with central charge $c=7/5$. 

The constructed $so(N)_1$ spin chains contain up to $N$-spin operators whose couplings are tuned to criticality by construction (namely, they are all are equal). However, exact solvability enables one to explore their phase diagrams beyond the critical point. By varying the couplings of nearest neighbor 2-spin with respect to $N$-spin terms, we showed that the phase diagrams of all odd $N$ models are qualitatively similar to the one of the simplest member of the hierarchy, the transverse field Ising chain, with the longer range couplings playing the role of the Zeeman term. For the even $N$ models the different terms could always be grouped to varying range generalizations of the XX- and YY-terms appearing in the XY model. Varying their relative couplings produced a phase diagram qualitatively similar to this canonical spin chain. While these results illustrate the general features of the phase diagrams, the number of terms in our $so(N)_1$ models grows linearly with $N$. For generic couplings the phase diagrams are expected to exhibit additional phases and critical points. We leave their study for future work. 

In addition to the full phase diagrams of our hierarchy of models, there are also several other interesting aspects that deserve further investigation. First, one
can use the exact solutions of the models to study the nature of the ground states, including correlations in them\cite{orus11} and their entanglement properties.\cite{Vidal03} In this respect it would be particularly interesting study the overlap of our $N$-local ground states with the projected Pfaffian states that are exact ground states of infinite range models with $so(N)_1$ criticality.\cite{Tu13} The low-energy theory of our models could also be studied and compared to that for the $so(N)$ symmetric spin chains.\cite{tu11} A second natural future direction is to consider various generalizations of our hierarchy models. The simplest generalization is to include a local magnetic field term (which is already present
in the TFI model). Even in the presence of such a term, all the models can still be solved
by means of a Jordan-Wigner transformation. On the other hand, if one adds a term
$\sigma^z_i \sigma^z_{i+1}$ to the critical XY chain, one obtains the so called
XXZ chain, which can be solved by the Bethe Ansatz\cite{bethe31}. It would be most interesting to
investigate if one can add analogous terms to our models and obtain
spin chains that allow for a solution via Bethe Ansatz techniques.

A third open question is the physical realization of our hierarchy models. As the locality of our models grows linearly with increasing $N$, the experimentally relevant ones reside in the small $N$ end of the hierarchy. Indeed, three-spin interactions required for the $N=3$ chain have been constructed in triangular optical lattices\cite{Pachos04}, but a challenge remains to isolate only those required for our hierarchy of models. Another way around the problem of many-spin interactions would be to treat the fermions as fundamental degrees of freedom. A proposal to implement the fermionic version of the TFI chain with cold atoms in an optical lattice has been put forward in Ref.~\onlinecite{Diehl11}. Since all our models are equivalent to free fermion problems of up to range $N$ tunneling and pairing, the realization of the $N=3$ chain with its interesting $su(2)_2$ critical point, could be possible along the same lines.

When viewed in terms of fermions, the odd $N$ models also generalize Kitaev's celebrated Majorana chain\cite{Kitaev01} that can be obtained from the TFI chain (which is the simplest odd $N$ model in our hierarchy). Unlike in two dimensions, in one spatial dimension symmetry class D, the symmetry class of the odd $N$ models, admits only two topologically distinct phases.\cite{Kitaev09, Schnyder09} In the phase diagrams we studied, these correspond to the 'spins aligned' phase with two-fold degenerate ground state (the analogue of the topological phase with Majorana end states for open boundary conditions) and the 'spin polarized' phase with unique ground state. As the critical point of our models corresponds to the transition between these two phases, the $so(N)_1$ criticality would usually imply that the adjacent phases would be somehow distinct from those with a critical point in the Ising universality class. Thus it would be fundamentally interesting to study whether there is any observable microscopic signature that distinguishes the gapped analogues of the 'topological' phases for $N=3,5,\ldots$ from those of the $N=1$ case. 

Finally, the application of the condensation picture\cite{Mansson13} to spin chains is far from complete. This is vividly illustrated by the presented spin-1 Blume-Capel example, which we strongly believe is just a single example of another hierarchy, exactly like the mapping between two TFI chains and the XY chains was just the simplest example of the $so(N)_1$ hierarchy. The fact that our construction works even in the absence of clear correspondence between the boundary conditions and the symmetry sectors suggests that there is something more fundamental to be understood about the form of condensing boundary term. A clue to this might be the way the constraining of boundary conditions closely resembles a Gutzwiller projection that has been used to construct $so(N)_1$ critical states.\cite{Tu13} Given such understanding, it would be interesting to generalize our method first to $N$ Blume-Capel models and then to other (higher) spin chains. This should also be possible in models whose criticality is not described by a CFT that is a direct product. Within the set of models we constructed, the counterpart of condensation could in principle be performed when $N$ is a multiple of 16, because then the CFT describing the criticality contains always a bosonic field. In Appendix~\ref{sec:characters} we comment on the possibility of using the condensation picture to obtain spin chains with critical points described by $(E_8)_1$, and products thereof. While such models are likely to have little experimental relevance, it would be academically highly satisfying to find a realization of such an exotic mathematical structure in terms of a spin chain.

{\em Acknowledgements --}
We thank Z.~Nussinov and G.~Sierra for interesting discussions.
We acknowledge financial support by the Swedish science research council
(TM, EA) and the Dutch Science Foundation NWO/FOM (VL).

\appendix

\section{The generalized condensing boundary term }
\label{app:boundary-hamiltonians}

In this appendix, we first explicitly derive the condensing boundary term \rf{HB3} for the system of three decoupled TFI chains. Then we give the most general form for these terms that takes into account possible additional signs that can appear for different chain lengths $L$.

\subsection{Derivation}

Let us consider a system of three decoupled critical TFI chains (we label them 0,1 and 2) described by $H^3_{\rm TFI}$ and consider condensing a boson in two of them, say chains 0 and 1, by adding the condensing boundary term
\be \label{HB01}
 H_{B,01}^2 = \bigl( \mathcal{P}_1 - \id \bigr) \sigma^{x}_{L-3}\sigma^{x}_{0} +
 \bigl( \mathcal{P}_0 - \id \bigr) \sigma^{x}_{L-2}\sigma^{x}_{1}\,.
\ee
By employing the duality transformations \rf{Transform2}, the resulting system $H^3_{\rm TFI}+H_{B,01}^2$ could be written as a decoupled system of a critical XY chain and a critical TFI chain. The criticality of this theory is described by $so(2)_1 \times$ Ising CFT, which also contains a boson. Recalling Example 2 in Section~\ref{sec:cond-ex}, condensation of this boson would lead to confinement of all the product primary fields that together with the $1$ and $\psi$ fields contain only a single $\sigma$, $\lambda_1$ or $\lambda_2$ field. These fields label the states in the total odd parity sectors  of $H^3_{\rm TFI}+H_{B,01}^2$, i.e. when the TFI chain and the XY chain have different boundary conditions. Defining $\mathcal{T}^z=\mathcal{P}_0 \mathcal{P}_1$ as the parity operator for the XY chain part of the system, we can force them to have always same boundary conditions, thus effectively condensing the boson in th $so(2)_1 \times$ Ising system, by adding to $H^3_{\rm TFI}+H_{B,01}^2$ the further condensing boundary term
\begin{equation}
\label{HBXY}
	H_B' = (\mathcal{T}^z-\id)\sigma^x_{L-1}\sigma^x_2+ (\mathcal{P}_2-\id)(\tau^x_{L-2}\tau^x_0+\tau^y_{L-2}\tau^y_0).
\end{equation}
The duality transformations \rf{Transform2} imply that $\tau^x_{L-2}\tau^x_0 = \mathcal{P}_1 \sigma_{L-3}^x \sigma_{0}^x$ and $\tau^y_{L-2}\tau^y_0 = \mathcal{P}_0 \sigma_{L-2}^x \sigma_{1}^x$. Thus the total condensing boundary term added to $H^3_{TFI}$ can be written as
\bq
 H^2_{B,01} + H_B' & = & (\mathcal{P}_0\mathcal{P}_1-\id)\sigma^x_{L-1}\sigma^x_2 + (\mathcal{P}_1 \mathcal{P}_2-\id)\sigma^x_{L-3}\sigma^x_0 \nonumber \\
  \ & \ & + (\mathcal{P}_0 \mathcal{P}_2-\id) \sigma^x_{L-2}\sigma^x_1.
\eq
The two boundary terms implementing the counterparts of the two condensations in critical spin chains are therefore equivalent to a single condensing boundary term. This term constrains the boundary conditions of the decoupled system such that all the three TFI chains can only have simultaneously periodic or anti-periodic boundary conditions across all the eight symmetry sectors labeled by $(\mathcal{P}_0,\mathcal{P}_1,\mathcal{P}_2)$. 

Since any sequence of condensing all the bosons in a system of $N$ decoupled TFI chains can be understood in terms of pairwise condensations of either this type ($N$ odd theory with $N'$ even theory) or that considered in the main text in Section~\ref{sec:boundaryterm}, any such process must be equivalent to adding the generalized boundary term \rf{HBN}.

\subsection{The most general $L$-dependent form}

In the main text, we restricted ourselves to the cases where the system size is an {\em even} multiple
of $N$. The reason we did this is that in the case that $L$ is an odd multiple of $N$, the
boundary Hamiltonians contain some additional signs. While they not important from the point of view of the condensation framework, they are required for the general duality transformations we present in Appendix~\ref{app:general-transform} to bring the $H_{so(N)_1}$ Hamiltonians into a translationally invariant form. In the end of the day, the additional signs come from the fact Pauli matrices obey the relation $\tau^{x} \tau^{y} = i \tau^{z}$, which when applied
an even, but not a multiple of four times, gives rise to a sign.

In the case that $N$ is even, the condensing boundary term takes the following
form
\begin{align}
H^{N}_{\rm B} &=
\sum_{n = 0}^{N-1}
\Bigl( \mathcal{S} \bigl( \prod_{\substack{l = 0\\  l\neq n}}^{N-1} P_{l} \bigr) -\id \Bigr)
\sigma^{x}_{L-N+n} \sigma^{x}_{n} \\
\nonumber
\mathcal{S} &= \begin{cases}
(-1)^{L/N} & \qquad \text{for } N \bmod 4 = 0 \\
1 & \qquad \text{for } N \bmod 4 = 2 \ ,
\end{cases}
\end{align}
while for case of odd $N$ they are given by 
\begin{align}
H_{B}^N &= \bigl( (\prod_{l = 1}^{N-1} P_{l}) - \id\bigr) \sigma^x_{L-N} \sigma^{x}_{0}
\\ \nonumber &
 + \sum_{n = 1}^{N-1}
\Bigl((-1)^{L/N} \bigl(\prod_{\substack{l = 0\\ l\neq n}}^{N-1} P_{l}\bigr) - \id \Bigr)
\sigma^x_{L-N + n} \sigma^{x}_{n} \ .
\end{align}
When $L$ an even multiple of $N$, both cases reduce to the form given in the main text as Eq.~\eqref{HBN}.

\section{Explicit forms of spin transformations}
\label{app:general-transform}

In this appendix we give the spin transformations that are necessary to bring the
Hamiltonians $H_{so(N)_1} = H^N_{\rm TFI} + H_{B}^N$ into a from that is manifestly translationally invariant.

The form of these transformations for general $N$ are rather unwieldy when expressed
completely in terms of the Pauli matrices $\tau$. To simplify the notation, we introduce a
set of string operators, that are closely related to the parity operators
$\mathcal{P}_n$ of Eq.~\eqref{eq:Pparity}, whose definition we repeat here
for convenience
\be
  \mathcal{P}_n = \prod_{j=0}^{L/N-1} \sigma^z_{jN+n}, \qquad n=0,1,\ldots,N-1.
\ee
The string operators we need are
\bq
  P_n^{<j} = \prod_{i<j} \sigma^z_{Ni+n}, \qquad  P_n^{>j} = \prod_{i>j} \sigma^z_{Ni+n},
\eq
where $n= 0,1,\ldots N-1$, and we have the relation $\mathcal{P}_n=P_n^{<j} \sigma^{z}_j P_n^{>j}$. Finally, we need the products of these operators
over all $n$, so we introduce
\begin{align}
P^{< j}_{t} &= \prod_{n = 0}^{N-1} P^{< j}_{n} &
P^{> j}_{t} &= \prod_{n = 0}^{N-1} P^{> j}_{n} \ .
\end{align}
Using this notation, the transformations for $\sigma^x$
in the case $N=4$ in Eq.~\eqref{Transform4} take the following form
\begin{align}
\sigma^{x}_{4j} &=
\tau^{y}_{4j}
P^{< j}_{t} P^{< j}_{0} \mathcal{P}_{1}
\\ \nonumber
\sigma^{x}_{4j+1} &=
\tau^{z}_{4j} \tau^{x}_{4j+1}
P^{< j}_{t} P^{< j}_{1}
\\ \nonumber
\sigma^{x}_{4j+2} &=
\tau^{y}_{4j+2} \tau^{z}_{4j+3}
P^{> j}_{t} P^{> j}_{2}
\\ \nonumber
\sigma^{x}_{4j+3} &=
\tau^{x}_{4j+3}
P^{> j}_{t} P^{> j}_{3} \mathcal{P}_{2} \ .
\end{align}
The generalization for arbitrary even $N$ reads as follows
\begin{widetext}
\begin{align}
\sigma^{z}_{jN} &= \tau^{x}_{jN} \tau^{z}_{jN+1} \cdots \tau^{z}_{jN+N-2} \tau^{x}_{jN+N-1}
&
\sigma^{x}_{jN} &= \tau^{y}_{jN}
P^{< j}_{t} P^{< j}_{0}
\bigl( \prod_{n = 1}^{N/2-1} \mathcal{P}_{n} \bigr)
\\ \nonumber
\sigma^{z}_{jN+1} &= \tau^{y}_{jN+1} \tau^{z}_{jN+2} \cdots \tau^{z}_{jN+N-3} \tau^{y}_{jN+N-2}
&
\sigma^{x}_{jN+1} &= \tau^{z}_{jN}\tau^{x}_{jN+1}
P^{< j}_{t} P^{< j}_{1}
\bigl( \prod_{n = 2}^{N/2-1} \mathcal{P}_{n} \bigr)
\\ \nonumber
&\vdots & &\vdots
\\ \nonumber
\sigma^{z}_{jN+N/2-2} &=
\tau^{x}_{jN+N/2-2} \tau^{z}_{jN+N/2-1} \tau^z_{jN+N/2} \tau^{x}_{jN+N/2+1}
&
\sigma^{x}_{jN+N/2-2} &=
\tau^{z}_{jN}\cdots \tau^{z}_{jN+N/2-3} \tau^{y}_{jN+N/2-2}
P^{< j}_{t} P^{< j}_{N/2-2}
\mathcal{P}_{N/2-1}
\\ \nonumber
\sigma^{z}_{jN+N/2-1} &= \tau^{y}_{jN+N/2-1} \tau^{y}_{jN+N/2}
&
\sigma^{x}_{jN+N/2-1} &=
\tau^{z}_{jN}\cdots \tau^{z}_{jN+N/2-2} \tau^{x}_{jN+N/2-1}
P^{< j}_{t} P^{< j}_{N/2-1}
\\ \nonumber
\sigma^{z}_{jN+N/2} &= \tau^{x}_{jN+N/2-1} \tau^{x}_{jN+N/2}
&
\sigma^{x}_{jN+N/2} &=
\tau^{y}_{jN+N/2} \tau^{z}_{jN+N/2+1}\cdots \tau^{z}_{jN+N-1} 
P^{> j}_{t} P^{> j}_{N/2}
\\ \nonumber
\sigma^{z}_{jN+N/2+1} &=
\tau^{y}_{jN+N/2-2} \tau^{z}_{jN+N/2-1} \tau^z_{jN+N/2} \tau^{y}_{jN+N/2+1}
&
\sigma^{x}_{jN+N/2+1} &=
\tau^{x}_{jN+N/2+1} \tau^{z}_{jN+N/2+2}\cdots \tau^{z}_{jN+N-1}
P^{> j}_{t} P^{> j}_{N/2+1}
\mathcal{P}_{N/2}
\\ \nonumber
&\vdots & &\vdots
\\ \nonumber
\sigma^{z}_{jN+N-2} &= \tau^{x}_{jN+1} \tau^{z}_{jN+2} \cdots \tau^{z}_{jN+N-3} \tau^{x}_{jN+N-2}
&
\sigma^{x}_{jN+N-2} &= \tau^{y}_{jN+N-2} \tau^{z}_{jN+N-1}
P^{> j}_{t} P^{> j}_{N-2}
\bigl( \prod_{n = N/2}^{N-3} \mathcal{P}_{n} \bigr)
\\ \nonumber
\sigma^{z}_{jN+N-1} &= \tau^{y}_{jN} \tau^{z}_{jN+1} \cdots \tau^{z}_{jN+N-2} \tau^{y}_{jN+N-1}
&
\sigma^{x}_{jN+N-1} &= \tau^{x}_{jN+N-1}
P^{> j}_{t} P^{> j}_{N-1}
\bigl( \prod_{n = N/2}^{N-2} \mathcal{P}_{n} \bigr)
\end{align}

When $N$ is odd, the transformations for the matrices $\sigma^z$ are given by
\begin{align}
\sigma^{z}_{jN} &=
\tau^{y}_{jN} \tau^{z}_{jN+1} \cdots
\tau^{z}_{jN+N-2} \tau^{y}_{jN+N-1}
\\ \nonumber
\sigma^{z}_{jN+1} &= \tau^{x}_{jN} \tau^{y}_{jN+1}
\\ \nonumber
\sigma^{z}_{jN+2} &= \tau^{x}_{jN+2} \tau^{y}_{jN+3}
\\ \nonumber
& \vdots
\\ \nonumber
\sigma^{z}_{jN+(N-1)/2-1} &= \tau^{x}_{jN+N-5} \tau^{y}_{jN+N-4}
\\ \nonumber
\sigma^{z}_{jN+(N-1)/2} &= \tau^{x}_{jN+N-3} \tau^{y}_{jN+N-2}
\\ \nonumber
\sigma^{z}_{jN+(N-1)/2+1} &= \tau^{y}_{jN+1} \tau^{x}_{jN+2}
\\ \nonumber
\sigma^{z}_{jN+(N-1)/2+2} &= \tau^{y}_{jN+3} \tau^{x}_{jN+4}
\\ \nonumber
&\vdots
\\ \nonumber
\sigma^{z}_{jN+N-2} &= \tau^{y}_{jN+N-4} \tau^{x}_{jN+N-3}
\\ \nonumber
\sigma^{z}_{jN+N-1} &= \tau^{y}_{jN+N-2} \tau^{x}_{jN+N-1}
\end{align}
Finally, the matrices $\sigma^x$ transform as
\begin{align}
\sigma^{x}_{jN} &=
\tau^{y}_{jN} \tau^{x}_{jN+1} \cdots
\tau^{y}_{jN+N-3} \tau^{x}_{jN+N-2} \tau^{y}_{jN+N-1}
P^{< j}_{t} \bigl(\prod_{n = 1}^{(N-1)/2} P^{< j}_{n} \bigr)
P^{> j}_{t} \bigl(\prod_{n = (N-1)/2+1}^{(N-1)} P^{< j}_{n} \bigr)
\\ \nonumber
\sigma^{x}_{jN+1} &= \tau^{y}_{jN}
P^{< j}_{t} P^{< j}_{1}
\bigl( \prod_{n = 2}^{(N-1)/2} \mathcal{P}_{n} \bigr)
\\ \nonumber
\sigma^{x}_{jN+2} &= \tau^{z}_{jN} \tau^{z}_{jN+1} \tau^{y}_{jN+2}
P^{< j}_{t} P^{< j}_{2}
\bigl( \prod_{n = 3}^{(N-1)/2} \mathcal{P}_{n} \bigr)
\\ \nonumber
&\vdots
\\ \nonumber
\sigma^{x}_{jN+(N-1)/2-1} &=
\tau^{z}_{jN} \cdots \tau^{z}_{jN+N-6} \tau^{y}_{jN+N-5}
P^{< j}_{t} P^{< j}_{(N-1)/2-1}
\mathcal{P}_{(N-1)/2}
\\ \nonumber
\sigma^{x}_{jN+(N-1)/2} &=
\tau^{z}_{jN} \cdots \tau^{z}_{jN+N-4} \tau^{y}_{jN+N-3}
P^{< j}_{t} P^{< j}_{(N-1)/2}
\\ \nonumber
\sigma^{x}_{jN+(N-1)/2+1} &=
\tau^{y}_{jN+2} \tau^{z}_{jN+3} \cdots \tau^{z}_{jN+N-1}
P^{> j}_{t} P^{> j}_{(N-1)/2+1}
\\ \nonumber
\sigma^{x}_{jN+(N-1)/2+2} &=
\tau^{y}_{jN+4} \tau^{z}_{jN+5} \cdots \tau^{z}_{jN+N-1}
P^{> j}_{t} P^{> j}_{(N-1)/2+2}
\mathcal{P}_{(N-1)/2+1}
\\ \nonumber
&\vdots
\\ \nonumber
\sigma^{x}_{jN+N-2} &= \tau^{y}_{jN+N-3} \tau^{z}_{jN+N-2} \tau^{z}_{jN+N-1}
P^{> j}_{t} P^{> j}_{N-2}
\bigl( \prod_{n = (N-1)/2+1}^{N-3} \mathcal{P}_{n} \bigr)
\\ \nonumber
\sigma^{x}_{jN+N-1} &= \tau^{y}_{jN+N-1}
P^{> j}_{t} P^{> j}_{N-1}
\bigl( \prod_{n = (N-1)/2+1}^{N-2} \mathcal{P}_{n} \bigr)
\end{align}
\end{widetext}

\section{The CFT predictions for the critical spectra}
\label{sec:cfttospectrum}

Conformal field theory gives a detailed prediction for the spectra of one-dimensional critical
systems. We refer to Ref.~\onlinecite{byb} for a general introduction to CFT.
Once the correct CFT for a given critical system has been identified, one can obtain
the spectrum in the thermodynamic limit. In particular, the energies of the states
of an $L$ site chain are given by
\begin{equation}
E = E_1 L - \frac{\pi v c}{6L} + \frac{2\pi v}{L} \bigl( h_{l} + h_{r} + n_{l} + n_{r} \bigr) \ .
\end{equation}
Here, the one-site energy $E_1$ and the velocity $v$ are non-universal numbers, while
the central charge $c$ and the scaling dimensions $h_{l}$ and $h_{r}$
can take several values, one for each primary field, and are determined
by the conformal field theory.
Finally, $n_{l}$ and $n_{r}$ are non-negative integers. The subscripts $l$ and $r$ refer
to the left and right moving modes of the CFT, which are decoupled.

To confirm that a particular finite size spectrum is conformal, one typically shifts the energy
of the states, such that the ground state has zero energy, $E_{\rm gs} = 0$.
In addition, one rescales the energies, such that the lowest excited state has energy
$E_{\rm ex,1} = h_{l} + h_{r}$. After this shift and rescaling, the spectrum is fixed completely,
and one can compare it to the spectrum predicted by CFT. The CFT spectrum takes the
form $E = h_{l} + h_{r} + n_{l} + n_{r} = 2h+ n_{l} + n_{r}$, were we assumed that the
left and right scaling dimensions are equal $h_{l} = h_{r} = h$, which will always be the
case for the theories we encounter in this paper. In addition, we also assumed that the
scaling dimension corresponding to the ground state is $h=0$.

CFT does not only predict which energies will be present in the spectrum, it also predicts
their degeneracies (and to some extent, their momenta). This information is encoded in the
partition function of the CFT. The total partition function splits into left and right moving pieces,
one for each primary field. In general, for a primary field $\phi_i$, with scaling dimension $h_i$,
the left moving part of the partition function reads
$Z_{l} (\phi_i) = q_{l}^{h_i} \sum_{n_l=0}^{\infty} c_{n_l} q^{n_l}_{l}$
(and similar for the right moving part),
where the $c_{n_l}$ are constants, depending on the primary field, and we
view $q_l$ as a formal variable. The total partition function
takes the form $Z_{\rm tot} = \sum_{i} Z_{l} (\phi_{i}) Z_{r} (\phi_{i})$, where the sum runs over all
primary fields in the theory. 

To explain how the partition function $Z_{\rm tot}$ encodes the energies of the states in the
spectrum of a critical model, we look at the `vacuum sector' of the CFT, which corresponds to the
trivial primary field, with scaling dimension $h=0$. The total partition function for this sector
takes the form
\begin{equation}
\begin{split}
& Z_{l} (\id) Z_{r} (\id) = \\ &1 + c_{1,0} q_l + c_{0,1} q_r  + c_{2,0} q^2_l
+ c_{1,1} q_l q_r + c_{0,2} q^2_r + \cdots
\end{split}
\end{equation}
Each term correspond to $c_{n_l,n_r}$ states, which have the energy $n_l+n_r$ (or, in general,
$2h + n_l + n_r$) that appears as the power of the $q_l$ variables. So, the degeneracy of the states is encoded in the constants $c_{n_l,n_r}$.

CFT does not completely predict the momenta of the states, but one can make the following
remarks. Let us assume that the state corresponding to the primary field $\phi_i$ has momentum
$k$ (in units of $2\pi /L$). This momentum is not fixed by the CFT. Often, but not always, the
momenta of the states obtained from this primary field by increasing the values $n_l$
and $n_r$ is given by $k-n_l + n_r$. This is typically true for Virasoro minimal models (such as
the Ising CFT). In the presence of additional symmetries, additional shifts in momenta can
occur (typically, shifts by $\pi$ or $\pi/2$)  upon increasing the values of $n_l$ and $n_r$. Thus,
one can not completely predict the momenta of all the states, even if the momenta of the states
corresponding to the primary fields are known.

In Appendix \ref{sec:characters} we explicitly state the partition functions for the $so(N)_1$ CFTs that are relevant to our hierarchy of spin models.

\section{Characters of the $so(N)_1$ CFTs}
\label{sec:characters}

In this appendix we will give the precise forms of the partition functions of the $so(N)_1$
CFTs describing the critical behavior of the hierarchy of spin chains we constructed. In the CFT literature, 
the partition functions $Z_{l}$ and $Z_{r}$ are often referred to as (chiral) characters of the
CFT, and denoted as $Z_{l} (\phi_i) = \ch_{q_l} (\phi_i)$ and $Z_{r} (\phi_i) = \ch_{q_r} (\phi_i)$. In this appendix we will adopt this notation.

The characters of $so(N)_1$ CFT were considered in, for instance, Ref.~\onlinecite{bs99}, using a so-called spinon formulation. Here we give the characters,
making use of the knowledge that they can be written in terms of $N$ free fermions, by employing the condensation picture.

\subsection{The characters for $N=1$}

For $N=1$, the $so(N)_1$ theory is just the Ising CFT, i.e. the minimal model with
central charge $c=\frac{1}{2}$, see Ref. \onlinecite{bpz84}.
Because the formulation of the $so(N)_1$
characters we use is based on the characters of the Ising theory, we give them here explicitly. 

We start by introducing the following notation, $(q)_m = \prod_{k=1}^{m} (1-q^k)$,
for $m\geq 1$ an integer. In addition, we define $(q)_0 = 1$ and
$(q)_\infty = \prod_{k=1}^{\infty} (1-q^k)$.
With this notation, we can write the (chiral) characters of the vacuum sector $\id$, the
$\sigma$-sector and the $\psi$-sector as follows
\begin{align}
\label{eq:chising}
\ch_{\id} (q) &= \sum_{\substack{m\geq 0 \\ {\rm even}}} \frac{q^{\frac{m^2}{2}}}{(q)_m} =
1 + q^2 + q^3 + 2 q^4 + 2 q^5 + \cdots \\ \nonumber
\ch_{\sigma} (q) &= q^{\frac{1}{16}}
\sum_{\substack{m\geq 0 \\ {\rm even}}} \frac{q^{\frac{m(m-1)}{2}}}{(q)_m} = 
\\ \nonumber &
q^{\frac{1}{16}} \bigl( 1 + q +  q^2 + 2 q^3 + 2 q^4 + 3 q^5 + \cdots \bigr) 
\\ \nonumber
\ch_{\psi} (q) &= \sum_{\substack{m\geq 1 \\ {\rm odd}}} \frac{q^{\frac{m^2}{2}}}{(q)_m} =
\\ \nonumber &
q^{\frac{1}{2}} \bigl( 1 + q +  q^2 + q^3 + 2 q^4 + 2 q^5 + \cdots \bigr) \ .
\end{align}

\subsection{The characters for $N=2$}
The case $so(2)_1$ is equivalent to a compactified free boson CFT, namely $u(1)_4$,
which has four primary fields. More detailed information on these CFTs can be found, for
instance, in Ref.~\onlinecite{fsz87}. 

In the general case $u(1)_p$, with $p$ an integer,
the fields are labeled by an integer $l=0,1,\ldots,p-1$, and the associated characters
read
\begin{equation}
\label{eq:u1character}
\ch_{p,l} (q) = \frac{1}{(q)_\infty}
\sum_{\substack{m \in \mathbb{Z}\\ m \bmod p = l}} q^{\frac{m^2}{2p}} \ .
\end{equation} 
We labeled the fields of the $so(2)_1$ theory as $\id$, $\lambda_1$, $\lambda_2$ and
$\psi$, which correspond to the labels $l=0$, $l=1$, $l=3$ and $l=2$, respectively. 
Using Eq.~\ref{eq:u1character}, we find the following results
\begin{align}
\ch^{N=2}_{\id} (q) &= \ch_{4,0} (q) =
\frac{1}{(q)_\infty}
\sum_{\substack{m \in \mathbb{Z}\\ m \bmod 4 = 0}} q^{\frac{m^2}{8}} =
\\ \nonumber &
1 + q + 4 q^2 + 5 q^3 + 9 q^4 + 13 q^5 + \cdots \\ \nonumber
\ch^{N=2}_{\lambda_1} (q) &= \ch^{N=2}_{\lambda_2} (q) = 
\ch_{4,1} (q) = \ch_{4,3} (q) =
\\ \nonumber &
\frac{1}{(q)_\infty}
\sum_{\substack{m \in \mathbb{Z}\\ m \bmod 4 =  1}} q^{\frac{m^2}{8}} = 
\\ \nonumber &
q^{\frac{1}{8}} \bigl( 1 + 2 q + 3 q^2 + 6 q^3 + 9 q^4 + 14 q^5 + \cdots \bigr) \\ \nonumber
\ch^{N=2}_{\psi} (q) &=
\ch_{4,2} (q) =
\frac{1}{(q)_\infty}
\sum_{\substack{m \in \mathbb{Z}\\ m \bmod 4 =  2}} q^{\frac{m^2}{8}} = 
\\ \nonumber &
q^{\frac{1}{2}} \bigl( 2 + 2 q + 4 q^2 + 6 q^3 + 12 q^4 + 16 q^5 + \cdots \bigr) \ .
\end{align}

\subsection{The characters for $N=3$}
With $N=3$, the model $so(3)_1$ is equivalent to $su(2)_2$. The general models
$su(2)_k$ have often been considered in the literature, see for instance the
Refs.~\onlinecite{book:lepowsky85,sf94,aks05},
which give details on the characters from rather different perspectives.
The characters take the following from (using the labels $\id,\sigma,\psi$)
\begin{align}
\ch^{N=3}_{\id} (q) &= \frac{1}{(q)_\infty}
\sum_{\substack{m_1 \geq 0 \\ m_2 \in \mathbb{Z}}}
\frac{q^{m_1^2 + 2 m_1 m_2 + m_2^2}}{(q)_{m_1}} =
\\ \nonumber &
1 + 3 q + 9 q^2 + 15 q^3 + 30 q^4 + 54 q^5 + \cdots \\ \nonumber
\ch^{N=3}_{\sigma} (q) &= \frac{q^{\frac{3}{16}}}{(q)_\infty}
\sum_{\substack{m_1 \geq 0 \\ m_2 \in \mathbb{Z}}}
\frac{q^{m_1^2 + 2 m_1 m_2 + m_2^2 + m_2}}{(q)_{m_1}} =
\\ \nonumber &
q^{\frac{3}{16}} \bigl( 2 + 6 q + 12 q^2 + 26 q^3 + 48 q^4 + 84 q^5 + \cdots \bigr) \\ \nonumber
\ch^{N=3}_{\psi} (q) &= \frac{q^{\frac{1}{2}}}{(q)_\infty}
\sum_{\substack{m_1 \geq 0 \\ m_2 \in \mathbb{Z}}}
\frac{q^{m_1^2 + 2 m_1 m_2 + m_2^2 + m_1 + 2 m_2}}{(q)_{m_1}} =
\\ \nonumber &
q^{\frac{1}{2}} \bigl( 3 + 4 q + 12 q^2 + 21 q^3 + 43 q^4 + 69 q^5 + \cdots \bigr) \ .
\end{align}

\subsection{The characters for $N=4$}
The model $so(4)_1$ is equivalent to $u(1)_2 \times u(1)_2$ (or
$su(2)_1 \times su(2)_1$), so the characters can be obtained
directly from Eq.~\eqref{eq:u1character}. Using the following
correspondence to the labels $(l_1,l_2)$ of the $u(1)_2 \times u(1)_2$ theory, 
$\id = (0,0)$, $\lambda_1 = (0,1)$,  $\lambda_2 = (1,0)$ and $\psi = (1,1)$, one finds
\begin{align}
\ch^{N=4}_{\id} (q) &=
1 + 6 q + 17 q^2 + 38 q^3 + 84 q^4 + 172 q^5 + \cdots \\ \nonumber
\ch^{N=4}_{\lambda_1} (q) &= \ch^{N=4}_{\lambda_2} (q) =
\\ \nonumber &
q^{\frac{1}{4}} \bigl( 2 + 8 q + 20 q^2 + 48 q^3 + 102 q^4 + 200 q^5 + \cdots \bigr) \\ \nonumber
\ch^{N=4}_{\psi} (q) &=  
\\ \nonumber &
q^{\frac{1}{2}} \bigl( 4 + 8 q + 28 q^2 + 56 q^3 + 124 q^4 + 232 q^5 + \cdots \bigr) \ .
\end{align}

\subsection{The characters for general $N$}
We now give the form of the characters for the general theory $so(N)_1$. In analogy to the condensation picture that we employed to construct the hierarchy of spin chains form $N$ decoupled TFI chains, these characters can be expressed in terms of the characters of the Ising model, namely
$\ch_{\id}(q)$, $\ch_{\sigma}(q)$ and $\ch_{\psi} (q)$, given in Eq.~\eqref{eq:chising}.

In the case $N=2$, one finds
\begin{align}
\nonumber
\ch^{N=2}_{\id} (q) &=
\ch^{N=1}_{\id} (q) \; \ch^{N=1}_{\id} (q) + \ch^{N=1}_{\psi} (q) \; \ch^{N=1}_{\psi} (q) \\ \nonumber
\ch^{N=2}_{\lambda_1} (q) &= \ch^{N=2}_{\lambda_2} (q)  =
\ch^{N=1}_{\sigma} (q) \; \ch^{N=1}_{\sigma} (q) \\
\ch^{N=2}_{\psi} (q) &=
2 \; \ch^{N=1}_{\id} (q) \; \ch^{N=1}_{\psi} (q) \ .
\end{align}
By using this result, we obtain the characters for the case $N=3$ as
\begin{align}
\ch^{N=3}_{\id} (q) &=
\ch^{N=1}_{\id} (q) \; \ch^{N=1}_{\id} (q) \; \ch^{N=1}_{\id} (q) +
\\ \nonumber &
3 \; \ch^{N=1}_{\id} (q) \; \ch^{N=1}_{\psi} (q) \; \ch^{N=1}_{\psi} (q) \\ \nonumber
\ch^{N=3}_{\sigma} (q) &=
2 \; \ch^{N=1}_{\sigma} (q) \; \ch^{N=1}_{\sigma} (q) \; \ch^{N=1}_{\sigma} (q) \\ \nonumber
\ch^{N=3}_{\psi} (q) &=
3 \; \ch^{N=1}_{\id} (q) \; \ch^{N=1}_{\id} (q) \; \ch^{N=1}_{\psi} (q) + 
\\ \nonumber &
\ch^{N=1}_{\psi} (q) \; \ch^{N=1}_{\psi} (q) \; \ch^{N=1}_{\psi} (q) \ .
\end{align}
From these results, it is not hard to obtain the structure for the general case of $so(N)_1$. For $N$ even they are given by
\begin{align}
\ch^{N}_{\id} (q) &= \sum_{\substack{p=0\\ p \; {\rm even}}}^N
\binom{N}{p}
\Bigl( \ch^{N=1}_{\id} (q) \Bigr)^{N-p} \Bigl( \ch^{N=1}_{\psi} (q) \Bigr)^{p} \\ \nonumber
\ch^{N}_{\lambda_1} (q) &= \ch^{N}_{\lambda_1} (q) =
2^{\frac{N}{2} - 1} \Bigl( \ch^{N=1}_{\sigma} (q) \Bigr)^{N} \\ \nonumber
\ch^{N}_{\psi} (q) &= \sum_{\substack{p=1\\ p \; {\rm odd}}}^{N-1}
\binom{N}{p}
\Bigl( \ch^{N=1}_{\id} (q) \Bigr)^{N-p} \Bigl( \ch^{N=1}_{\psi} (q) \Bigr)^{p},
\end{align}
while for $N$ odd the characters read
\begin{align}
\ch^{N}_{\id} (q) &= \sum_{\substack{p=0\\ p \; {\rm even}}}^{N-1}
\binom{N}{p}
\Bigl( \ch^{N=1}_{\id} (q) \Bigr)^{N-p} \Bigl( \ch^{N=1}_{\psi} (q) \Bigr)^{p} \\ \nonumber
\ch^{N}_{\sigma} (q) &=
2^{\frac{N-1}{2}} \Bigl( \ch^{N=1}_{\sigma} (q) \Bigr)^{N} \\ \nonumber
\ch^{N}_{\psi} (q) &= \sum_{\substack{p=1\\ p \; {\rm odd}}}^{N}
\binom{N}{p}
\Bigl( \ch^{N=1}_{\id} (q) \Bigr)^{N-p} \Bigl( \ch^{N=1}_{\psi} (q) \Bigr)^{p}.
\end{align}

\subsection{Remark about the character of $(E_{8})_{1}$}

We close this appendix by making a remark about the CFT associated with
$(E_8)_1$. When $N=16$, the scaling dimensions
of the fields $\lambda_1$ and $\lambda_2$ of the $so(N)_1$ CFT are integers.
This means that one could
condense, for instance, the field $\lambda_1$. It turns out that after condensation, the
fields $\lambda_2$ and $\psi$ are confined, so one is left with a theory which consists
of only the vacuum sector, and has central charge $c=8$. This is the so-called $(E_8)_1$
CFT.

The character of this theory reads
\begin{equation}
\ch_{(E_8)_1} (q) = \ch^{N=16}_{\id} (q) + \ch^{N=16}_{\lambda_1} (q) \ .
\end{equation}
In principle, one could try to take the spin chain, which has $so(16)_1$ as its critical
behavior, and add a boundary term, which causes the condensation to the $(E_8)_1$
critical behavior. It turns out, however, that if one constructs the full spectrum associated
with the $(E_8)_1$ CFT, one obtains exactly the same energies as predicted by
the $so(16)_1$ CFT, because of the following relation
\begin{equation}
\begin{split}
\bigl(\ch_{(E_8)_1} (q)\bigr)^2 &= 
\bigl(\ch^{N=16}_{\id} (q)\bigr)^2 + 
\bigl(\ch^{N=16}_{\lambda_1} (q)\bigr)^2 + 
\\ &
\bigl(\ch^{N=16}_{\lambda_2} (q)\bigr)^2 + 
\bigl(\ch^{N=16}_{\psi} (q)\bigr)^2 \ .
\end{split}
\end{equation}
In principle, the momenta of the states could differ, but because
CFT does not fully specify the precise momenta, one can not unambiguously say if a
certain spectrum is described by the $so(16)_1$ or $(E_8)_1$ CFT.
It is more a matter of choice how one interprets the spectrum.

In the case that $N=32$, the situation is slightly different. The theory $so(32)_1$ also
contains two bosons, this time with $h_\lambda = 2$, which can be added to the chiral
algebra. One obtains the theory $(E_8)_1 \times (E_8)_1$,
\begin{equation}
\ch_{(E_8)_1} (q) \times  \ch_{(E_8)_1} (q) =
\ch^{N=32}_{\id} (q) + \ch^{N=32}_{\lambda_1} (q) \ .
\end{equation}
However, in this case, the number of states in the spectra differs between the theories
$(E_8)_1 \times (E_8)_1$ and $so(32)_1$, because of the inequality
\begin{equation}
\begin{split}
\bigl(\ch_{(E_8)_1} (q)\bigr)^4 & \neq 
\bigl(\ch^{N=32}_{\id} (q)\bigr)^2 + 
\bigl(\ch^{N=32}_{\lambda_1} (q)\bigr)^2 +
\\ & 
\bigl(\ch^{N=32}_{\lambda_2} (q)\bigr)^2 + 
\bigl(\ch^{N=32}_{\psi} (q)\bigr)^2 \ .
\end{split}
\end{equation}
So, it should be possible to take the spin chain with $so(32)_1$ critical behavior, and
add an appropriate boundary term, to obtain a critical spin chain described by the 
$(E_8)_1 \times (E_8)_1$ CFT. We did not embark on this exercise, however.

Finally, we note that for $N=16 p$, with $p$ an integer $p\geq 3$, the spectrum of
the theory $so(16 p)_1$ contains two bosons, with integer scaling dimension $p > 2$.
Adding one of these bosons does not directly give the $\bigl((E_8)_1\bigr)^p$ CFT
as one might have expected naively. Instead, one finds the following relations
for $N=48,64,80,96$
\begin{align}
& \ch^{N=48}_{\id} (q) + \ch^{N=48}_{\lambda_1} (q) =
\\ \nonumber &
\bigl(\ch_{(E_8)_1} (q)\bigr)^3 + 348 q \\ \nonumber 
& \ch^{N=64}_{\id} (q) + \ch^{N=64}_{\lambda_1} (q) =
\\ \nonumber &
\bigl(\ch_{(E_8)_1} (q)\bigr)^4 + 1024 q \; \ch_{(E_8)_1} (q) \\ \nonumber
& \ch^{N=80}_{\id} (q) + \ch^{N=80}_{\lambda_1} (q) =
\\ \nonumber &
\bigl(\ch_{(E_8)_1} (q)\bigr)^5 + 1920 q \; \bigl(\ch_{(E_8)_1} (q) \bigr)^2 \\ \nonumber
& \ch^{N=96}_{\id} (q) + \ch^{N=96}_{\lambda_1} (q) =
\\ \nonumber &
\bigl(\ch_{(E_8)_1} (q)\bigr)^6 + 3072 q \; \bigl(\ch_{(E_8)_1} (q) \bigr)^3 + 98304 q^2
\end{align}
In general, the sum 
$\ch^{N=16 p}_{\id} (q) + \ch^{N=16p}_{\lambda_1} (q)$
decomposes in terms of
$$\bigl(\ch_{(E_8)_1} (q)\bigr)^p \; , \bigl(\ch_{(E_8)_1} (q)\bigr)^{p-3} \; , \ldots,
\bigl(\ch_{(E_8)_1} (q)\bigr)^{p \bmod 3} \ .$$
For completeness, we give the explicit decomposition. For $p$ a positive integer,
we have
\begin{equation}
\begin{split}
& \ch^{N=16 p}_{\id} (q) + \ch^{N=16p}_{\lambda_1} (q) =
\\ &
\sum_{j=0}^{\lfloor \frac{p}{3}\rfloor}
2^{8j}\frac{p}{p - j}\binom{p-j}{2j} q^j \bigl(\ch_{(E_8)_1} (q)\bigr)^{p-3j} \ ,
\end{split}
\end{equation}
where the floor function $\lfloor x \rfloor$ denotes the largest integer $j$, such that
$j \leq x$.



\begin{thebibliography}{10}

\bibitem{bethe31}
H.~Bethe,
{\it Zur Theorie der Metalle. I. Eigenwerte und Eigenfunktionen der linearen Atomkette.}
Z. Phys. {\bf 71}, 205 (1931).

\bibitem{book:sachdev}
S.~Sachdev,
{\it Quantum phase transitions},
Cambridge University Press, Cambridge (1999).

\bibitem{epw70}
R.J.~Elliott, P.~Pfeuty, C.~Wood,
{\it Ising model with a transverse field},
Phys. Rev. Lett. {\bf 25}, 443 (1970).

\bibitem{p70}
P.~Pfeuty,
{\it The one-dimensional Ising model with a transverse field},
Ann. Phys. {\bf 57}, 79 (1970).

\bibitem{lsm61}
E.~Lieb, T.~Schultz, D.~Mattis,
{\it Two Soluble Models of an Antiferromagnetic Chain},
Ann. Phys. {\bf 16}, 407 (1961).

\bibitem{bpz84}
A.A.~Belavin, A.M.~Polyakov, A.B.~Zamolodchikov,
{\it Infinite conformal symmetry in two-dimensional
quantum field theory}, 
Nucl. Phys. B {\bf 241}, 333 (1984).

\bibitem{byb}
P.~Di~Francesco, P.~Mathieu, D.~S\'en\'echal,
{\it Conformal field theory},
Springer, New York (1999).

\bibitem{zamolodchkov80}
A.B.~Zamolodchikov, V.A.~Fateev,
{\it Model factorized S-matrix and an integrable spin-1 Heisenberg chain},
Sov. J. Phys. {\bf 32}, 298 (1980).

\bibitem{takhtajan82}
L.A.~Takhtajan,
{\it The picture of low-lying excitations in the isotropic Heisenberg chain of arbitrary spins},
Phys. Lett. A {\bf 87}, 479 (1982).

\bibitem{babujian83}
H.M.~Babujian,
{\it Exact solution of the isotropic Heisenberg chain with arbitrary spins:
Thermodynamics of the model},
Nucl. Phys. B {\bf 215}, 317 (1983).

\bibitem{haldane88}
F.D.M.~Haldane,
{\it Exact Jastrow-Gutzwiller resonating-valence-bond ground state of the
spin-1/2 antiferromagnetic Heisenberg chain with $1/r^2$ exchange},
Phys. Rev. Lett. {\bf 60}, 635 (1988).

\bibitem{shastry88}
S.B.~Shastry,
{\it Exact solution of an S=1/2 Heisenberg antiferromagnetic chain with long-ranged
interactions},
Phys. Rev. Lett. {\bf 60}, 639 (1988),

\bibitem{nielsen11}
A.E.B.~Nielsen, J.I.~Cirac, G.~Sierra,
{\it Quantum spin Hamiltonians for the $SU(2)_k$ WZW model},
J. Stat. Mech. P11014 (2011).

\bibitem{book:greiter}
M. Greiter, 
{\it Mapping of Parent Hamiltonians: From Abelian and non-Abelian Quantum Hall States to
Exact Models of Critical Spin Chains},
Springer Tracts in Modern Physics, Vol. {\bf 244} (2011).

\bibitem{michaud13}
F.~Michaud, S.R.~Manmana, F~Mila, 
{\it Realization of higher Wess-Zumino-Witten models in spin chains},
Phys. Rev. B {\bf 87}, 140404(R) (2013).

\bibitem{mg69a}
C.K.~Majumdar, D.K.~Ghosh,
{\it On Next-Nearest-Neighbor interaction in linear chain. I},
J. Math. Phys. {\bf 10}, 1388 (1969).

\bibitem{mg69b}
C.K.~Majumdar, D.K.~Ghosh,
{\it On Next-Nearest-Neighbor interaction in linear chain. II},
J. Math. Phys. {\bf 10}, 1399 (1969).

\bibitem{tu08}
H.-H.~Tu, G.-M.~Zhang,
{\it Class of exactly solvable $SO(n)$ symmetric spin chains with matrix product ground states},
Phys. Rev. B {\bf 78}, 094404 (2008). 

\bibitem{alet11}
F.~Alet, S.~Capponi, H.~Nonne, P.~Lecheminant, I.P.~McCulloch,
{\it Quantum criticality in the $SO(5)$ bilinear-biquadratic Heisenberg chain},
Phys. Rev. B {\bf 83}, 060407(R) (2011). 

\bibitem{tu11}
H.-H.~Tu, R.~Or\'{u}s
{\it Effective field theory for the $SO(n)$ bilinear-biquadratic spin chain},
Phys. Rev. Lett. {\bf }, 077204 (2011).

\bibitem{orus11}
R.~Or\'us, T.-C.~Wei, H.-H~Tu,
{\it Phase diagram of the SO(n) bilinear-biquadratic chain from many-body entanglement},
Phys. Rev B {\bf 84}, 064409 (2011).

\bibitem{capponi13}
S.~Capponi, P.~Lecheminant, M.~Moliner,
{\it Quantum phase transitions in multileg spin ladders with ring exchange},
Phys. Rev. B {\bf 88}, 075132 (2013).

\bibitem{Tu13}
H.-H.~Tu,
{\it Projected BCS states and spin Hamiltonians for the $so(n)_1$
Wess-Zumino-Witten model},
Phys. Rev. B {\bf 87}, 041103(R) (2013).

\bibitem{Mansson13}
T.~M\r{a}nsson, V.~Lahtinen, J.~Suorsa, E.~Ardonne,
{\it Condensate-induced transitions and critical spin chains},
Phys. Rev. B {\bf 88}, 041403(R) (2013).

\bibitem{Bais09}
F.A.~Bais, J.K.~Slingerland,
{\it Condensate-induced transitions between topologically ordered phases},
Phys. Rev. B {\bf 79}, 045316 (2009).

\bibitem{Witten89}
E.~Witten, 
{\it Quantum field theory and the Jones polynomial},
Comm. Math. Phys. {\bf 121}, 351 (1989).

\bibitem{suzuki71b}
M.~Suzuki,
{\it Relationship among Exactly Soluble Models of Critical Phenomena. I},
Progr. Theor. Phys. {\bf 46}, 1337 (1971).

\bibitem{Dzyaloshinskii58}
I.~Dzyaloshinsky
{\it A thermodynamic theory of ``weak'' ferromagnetism of antiferromagnetics},
J. Phys. Chem. Solids {\bf 4}, 241 (1958).

\bibitem{Moriya60}
T.~Moriya,
{\it Anisotropic superexchange interaction and weak ferromagnetism},
Phys. Rev. {\bf 120}, 91 (1960).

\bibitem{Pachos04}
J.~K.~Pachos and M.~B.~Plenio, 
{\it Three-spin interactions in optical lattices and criticality in cluster Hamiltonians},
Phys. Rev. Lett. {\bf 93}, 056402 (2004).

\bibitem{Witten84}
E.~Witten,
{\it Non-abelian bosonization in two dimensions},
Comm. Math. Phys. {\bf 92}, 455 (1984). 

\bibitem{blume66}
M.~Blume,
{\it Theory of the First-Order Magnetic Phase Change in $UO_2$},
Phys. Rev. {\bf 141}, 517 (1966).

\bibitem{capel66}
H.W.~Capel,
{\it On the possibility of first-order phase transitions in Ising systems of triplet ions
with zero-field splitting},
Physica {\bf 32}, 966 (1966).

\bibitem{k06}
A.~Kitaev,
{\it Anyons in an exactly solved model and beyond},
Ann. Phys. {\bf 321}, 2 (2006).

\bibitem{Read00}
N.~Read, D.~Green,
{\it Paired states of fermions in two dimensions with breaking of parity and time-reversal symmetries, and the fractional quantum Hall effect},
Phys. Rev. B {\bf 61}, 10267 (2000).

\bibitem{Lahtinen12}
V.~Lahtinen, A.W.W.~Ludwig, J.K.~Pachos, S.~Trebst,
{\it Topological liquid nucleation induced by vortex-vortex interactions in Kitaev's honeycomb model},
Phys. Rev. B {\bf 86}, 075115 (2012).

\bibitem{capel77}
H.W.~Capel, J.H.H.~Perk,
{\it Autocorrelation function of the x-component of the
magnetization in the one-dimensional XY-model},
Physica {\bf A} 87, 211 (1977).

\bibitem{jf78}
R.~Jullien, J.N.~Fields,
{\it Equivalence between a spin 1/2 dimerized XY chain and two
independent Ising chains in transverse fields},
Phys. Lett. A {\bf 69}, 214 (1978).

\bibitem{pjp82}
K.A.~Penson, R.~Jullien, P.~Pfeuty,
{\it Phase transitions in systems with multispin interactions},
Phys. Rev. B {\bf 26}, 6334 (1982).

\bibitem{f94}
D.S.~Fisher,
{\it Random antiferromagnetic quantum spin chains},
Phys. Rev. B {\bf 50}, 3799 (1994).

\bibitem{Gottlieb99}
D. Gottlieb and J. R\"{o}ssler,
{\it Exact solution of a spin chain with binary and ternary interactions of Dzialoshinsky-Moriya type},
Phys. Rev. B {\bf 60}, 9232 (1999).

\bibitem{Zvyagin06}
A. A. Zvyagin and G. A. Skorobagat'ko,
{\it Exactly solvable quantum spin model with alternating and multiple spin exchange interactions},
Phys. Rev. B {\bf 73}, 024427 (2006).

\bibitem{Kitaev01}
A.Y.~Kitaev,
{\it Unpaired Majorana fermions in quantum wires},
Phys. Usp. {\bf 44}, 131 (2001).

\bibitem{Kitaev09}
A.Y.~Kitaev,
{\it Periodic table for topological insulators and superconductors},
AIP Conf. Proc. {\bf 1134}, 22 (2009).

\bibitem{Schnyder09}
A.P.~Schnyder, S.~Ryu, A.~Furusaki, A.W.W.~Ludwig,
{\it Classification of Topological Insulators and Superconductors},
AIP Conf. Proc. {\bf 1134}, 10 (2009).

\bibitem{vongehlen90}
G.~von~Gehlen,
{\it Off-criticality behaviour of the Blume-Capel quantum chain as a check of
Zamolodchikov's conjecture},
Nucl. Phys. B {\bf 330}, 741 (1990).

\bibitem{kennedy92}
T.~Kennedy, H.~Tasaki,
{\it Hidden Symmetry Breaking and the Haldane Phase in $S = 1$
Quantum Spin Chains},
Comm. Math. Phys. {\bf 147}, 431 (1992).

\bibitem{oshikawa92}
M.~Oshikawa,
{\it Hidden $Z_2 \times Z_2$ symmetry in quantum spin chains with arbitrary integer spin},
J. Phys.: Condens. Matter {\bf 4}, 7469 (1992).

\bibitem{friedan84}
D.~Friedan, Z.~Qiu, S.~Shenker,
{\it Conformal invariance, unitarity, and critical exponents in two dimensions},
Phys. Rev. Lett. {\bf 52}, 1575 (1984).

\bibitem{cappelli87}
A.~Cappelli,
{\it Modular invariant partition functions of superconformal theories},
Phys. Lett. B {\bf 185}, 82 (1987).

\bibitem{r85}
A.~Rocha-Caridi,
{\it Vacuum vector representations of the Virasoro algebra},
in {\em Vertex Operators in Mathematics and Physics},
J.~Lepowski, S.~Mandelstam, J.~Singer, eds., p. 451-473,
Math. Sci. Res. Inst. Publ., no. 3, Springer-Verlag (1984).

\bibitem{gko86}
P.~Goddard, A.~Kent, D.~Olive,
{\it Unitary Representations of the Virasoro and Super-Virasoro Algebras},
Comm. Math. Phys. {\bf 103}, 105 (1986).

\bibitem{Vidal03}
G.~Vidal, J.~I.~Latorre, E.~Rico and A.~Kitaev,
{\it Entanglement in quantum critical phenomena},
Phys. Rev. Lett. {\bf 90}, 227902 (2003).


\bibitem{Diehl11}
S.~Diehl, E.~Rico, M.A.~Baranov, P.~Zoller,
{\it Topology by Dissipation in Atomic Quantum Wires},
Nature Physics {\bf 7}, 971 (2011).


\bibitem{bs99}
P.~Bouwknegt, K.~Schoutens,
{\it Exclusion statistics in conformal field theory - Generalized fermions and spinons
for level-1 WZW theories},
Nucl. Phys. B {\bf 547}, 501 (1999). 

\bibitem{fsz87}
P.~Di~Francesco, H.~Saleur, J.B.~Zuber,
{\it Modular invariance in non-minimal two-dimensional conformal theories},
Nucl. Phys. B {\bf 285}, 454 (1987).

\bibitem{book:lepowsky85}
J.~Lepowsky, M~Primc,
{\it Structure for Standard modules of the affine Lie algebra $A^{(1)}_1$},
Contemporary Mathematics,  Vol. {\bf 46} (Providence RI: AMS) (1985).

\bibitem{sf94}
A.V.~Stoyanovsky, B.L.~Feigin,
{\it Functional models for representations of current algebras and semi-infinite
Schubert cells},
Funct. Anal. Appl. {\bf 28}, 55 (1994).

\bibitem{aks05}
E.~Ardonne, R.~Kedem, M.~Stone,
{\it Filling the Bose sea: symmetric quantum Hall edge states and affine characters},
J. Phys. A {\bf 38}, 617 (2005).

\bibitem{footnote1}
We note that the operators $c_k$ are not directly the fourier transform
of the operators $c_j$, but obtained after an additional Bogoliubov
transformation.

\bibitem{footnote2}
In general, the Blume-Capel model also contains a term $\gamma (S_i^z)^2$,
but we only consider the case $\gamma=0$, because for small $\gamma$, the only
effect of this term is to change the location of the tri-critical point\cite{vongehlen90}.


\end{thebibliography}
\end{document}